\documentclass[11pt,a4paper]{article}
\usepackage{jinstpub}
\usepackage{bm}
\usepackage{booktabs}
\usepackage{array}
\usepackage{siunitx}
\usepackage{xcolor}
\usepackage{microtype}

\graphicspath{{./}}

\newcommand{\notretained}{\textit{Not retained in the archived campaign summary}}

\title{NTL amplification in a SuperCDMS-style detector fabricated from a University of South Dakota-grown germanium crystal}

\author[a,1]{Dongming~Mei,\note{Corresponding author.}}
\author[b]{Gerardo~Gonzalez,}
\author[b]{William~Baker,}
\author[b]{Mark~Platt}
\author[b]{and Nader~Mirabolfathi}

\affiliation[a]{Department of Physics, University of South Dakota,\\
414 E. Clark St., Vermillion, South Dakota 57069, USA}
\affiliation[b]{Department of Physics and Astronomy, Texas A\&M University,\\
College Station, Texas 77843, USA}

\emailAdd{dongming.mei@usd.edu}

\abstract{Neganov--Trofimov--Luke (NTL) amplification enables the eV--sub-keV thresholds needed for low-mass dark-matter and coherent elastic neutrino--nucleus scattering experiments. We tested two University of South Dakota (USD)-grown germanium crystals fabricated as SuperCDMS-style four-channel athermal-phonon detectors. Both detectors had nominal envelope dimensions of \SI{75}{\milli\meter} diameter and \SI{10}{\milli\meter} thickness, with reported masses of \SI{188}{\gram} and \SI{230}{\gram}, respectively, and were operated at an approximate bath temperature of \SI{10}{\milli\kelvin}. The crystals had (100) orientation, a net electrically active impurity concentration near $10^{10}~\mathrm{cm}^{-3}$, and an etch-pit density of $300$--$7000~\mathrm{cm}^{-2}$. Contact-free bias configurations showed no resolvable gain. For Detector~2, a continuous \SI{600}{\nano\meter} aluminum bias film was deposited with a modified Perkin Elmer 4400 sputter coater. In the contacted configuration, Detector~2 exhibited clear voltage-dependent NTL amplification. The nominal reconstructed channel-B absorbed energy increased from approximately \SI{3.9}{\kilo\electronvolt} at \SI{0}{\volt} to \SI{34.9}{\kilo\electronvolt} at \SI{40}{\volt}, corresponding to an enhancement factor of approximately 8.9. Raw-data reanalysis gives fit-returned statistical centroid uncertainties below 0.3\% across all seven bias settings. A low-bias phenomenological fit yielded effective prompt- and NTL-phonon response coefficients of $(6.77\pm0.82)\%$ and $(4.77\pm0.20)\%$, respectively; these quoted uncertainties are regression standard errors and do not include the full calibration, TES-response, or absorbed-energy systematic uncertainty. These results demonstrate proof-of-principle compatibility of USD-grown germanium with SuperCDMS-style high-voltage phonon detectors and show that material qualification must account for both crystal properties and detector fabrication.}


\begin{document}
\maketitle
\flushbottom

\section{Introduction}

Cryogenic semiconductor detectors convert small particle-induced energy deposits into measurable athermal phonons. When electron--hole pairs drift through an electric potential, the work performed by the field is released as additional phonons through the Neganov--Trofimov--Luke (NTL) effect \cite{Neganov1985,Luke1988}. For an electron recoil of energy $E_R$ in germanium, the ideal total phonon energy is
\begin{equation}
E_{\mathrm{ph}}=E_R+N_{eh}eV
=E_R\left(1+\frac{eV}{\varepsilon}\right),
\label{eq:ntl_gain}
\end{equation}
where $N_{eh}=E_R/\varepsilon$, $V$ is the potential traversed by the charge carriers, and $\varepsilon\simeq\SI{3.0}{\electronvolt}$ is the mean energy required to create one electron--hole pair in germanium. Equation~\eqref{eq:ntl_gain} shows why the technique is powerful: the phonon signal grows approximately linearly with traversed voltage while the baseline readout noise need not grow by the same factor.

This internal amplification directly addresses one of the central experimental challenges in low-mass dark-matter detection. Sub-GeV dark matter can produce only eV--keV-scale electronic or nuclear excitations, so sensitivity is controlled by the ability to resolve one or a few electron--hole pairs and the associated phonons. Voltage-assisted cryogenic detectors have consequently enabled leading low-mass searches in CDMSlite, SuperCDMS HVeV, and EDELWEISS \cite{Agnese2014,Agnese2016,Agnese2018CDMSlite,Agnese2018HVeV,Arnaud2020}. The planned SuperCDMS SNOLAB high-voltage program extends this strategy to larger exposures and motivates crystals with reproducible charge transport, low leakage, stable high-field operation, and robust phonon-sensor integration \cite{Agnese2017}.

A second strong motivation is coherent elastic neutrino--nucleus scattering (CE$\nu$NS), predicted by Freedman and first observed by the COHERENT Collaboration \cite{Freedman1974,Akimov2017}. Because reactor and solar neutrinos generate predominantly sub-keV nuclear recoils, lower threshold translates directly into increased CE$\nu$NS acceptance and sensitivity to neutrino properties and physics beyond the Standard Model. Germanium detectors have provided evidence and observations of reactor CE$\nu$NS at very low thresholds \cite{Colaresi2022,Ackermann2025}, while phonon-mediated high-voltage detector concepts explicitly target both low-mass dark matter and reactor CE$\nu$NS \cite{Neog2022}. NTL-amplified Ge detectors are therefore an enabling technology for two complementary frontier measurements.

An essential qualification principle follows from the NTL mechanism: observing gain depends on both the grown crystal and the fabricated detector. The crystal must provide sufficiently low electrically active impurity concentration, useful carrier transport, and limited bulk trapping. The fabricated device must establish a stable internal field, suppress uncontrolled carrier injection and surface leakage, maintain suitable electrode and surface boundary conditions, collect a useful fraction of the generated phonons, and operate its QET/TES channels with sufficiently low noise. High-voltage phonon-detector studies show that trapping, impact ionization, surface and electrode effects, leakage backgrounds, phonon partition, and sensor fabrication all shape the measured response \cite{Ponce2020,Albakry2022,Wilson2024,Neog2022,Ren2021}. Previous detectors fabricated from USD-grown crystals likewise demonstrated that contact preparation and device geometry can determine leakage current, depletion behavior, and spectroscopic performance even when the starting material is usable \cite{Wei2018,Meng2019,Panth2020}. A null NTL result in one device configuration therefore does not by itself disqualify the crystal; conversely, measurable gain in a functioning device is a direct functional test of the material--device combination.

Producing detector-grade high-purity germanium (HPGe) is unusually difficult. Centimeter-scale depletion requires a net electrically active impurity concentration typically in the $10^{9}$--$10^{10}\,\mathrm{cm}^{-3}$ range, together with controlled dislocation density, low concentrations of trapping centers, and sufficiently uniform radial and axial impurity profiles \cite{Haller1981,Abrosimov2020}. During Czochralski growth, these properties depend on coupled heat, momentum, and mass transport; impurity segregation; melt convection; interface shape; pull and rotation rates; gas environment; and contamination control. Small deviations can substantially reduce the fraction of a boule that is detector grade \cite{Wang2012,Wang2015USD,Bhattarai2024,Prem2026}. The complete chain also requires zone refining, high-temperature crystal pulling, Hall-effect and defect characterization, precision mechanical processing, surface treatment, contact fabrication, and low-noise detector testing.

Detector-grade HPGe production is therefore maintained by a small, specialized commercial and research community, and limited growth capacity remains a concern for future rare-event experiments \cite{Abrosimov2020,Mei2024,Prem2026}. USD operates a U.S. university-based program integrating germanium purification, Czochralski growth, crystal characterization, detector fabrication, and detector testing \cite{Wang2015USD,Wei2018,Meng2019,Mei2024}. This research-based capability enables rapid feedback between materials science and detector performance and provides a domestic training and technology-development platform complementary to commercial procurement.

This paper asks a deliberately limited system-level question: can a USD-grown crystal, after SuperCDMS-style fabrication and cryogenic integration, sustain charge drift through an applied potential and generate measurable NTL phonons? The answer is affirmative for the tested crystal in the final contacted configuration. The result demonstrates proof-of-principle compatibility with continued SuperCDMS-style high-voltage detector development. At the same time, incomplete channel yield, the absence of direct leakage-current monitoring, and nonuniform phonon collection identify the key detector-engineering priorities for the next development stage.

\section{Crystal provenance, device fabrication, and experimental campaign}
\label{sec:experiment}

\subsection{Scope of the material validation}

The campaign involved two USD-grown Ge substrates that were subsequently instrumented as four-channel athermal-phonon detectors. The archived campaign record gives nominal envelope dimensions of \SI{75}{\milli\meter} diameter and \SI{10}{\milli\meter} thickness for both substrates, with a (100) surface orientation; Detector~1 had a reported mass of \SI{188}{\gram} and Detector~2 a reported mass of \SI{230}{\gram}. The nominal dimensions should not be interpreted as a complete as-built solid-cylinder geometry, particularly for Detector~1: the detailed edge/bevel geometry, thickness profile, and records of any material removed during detector processing were not retained in the present record. We therefore report the measured masses and nominal envelope dimensions separately rather than using them to infer detector density or active volume. Hall-effect measurements indicated a net electrically active impurity concentration of approximately $10^{10}~\mathrm{cm}^{-3}$, near the upper end of the detector-grade range discussed in Sec.~I \cite{Haller1981,Abrosimov2020}. Etch-pit measurements yielded $300<\rho_{\mathrm{EPD}}<7000~\mathrm{cm}^{-2}$, providing an experimental measure of the dislocation-density range. These properties give direct material context for the functional NTL test. Both detector assemblies were operated at an approximate bath temperature of \SI{10}{\milli\kelvin}. The $^{241}$Am source activity was $1~\mu\mathrm{Ci}$; the source-to-detector distances were \SI{35.5}{\milli\meter} for Detector~1 and \SI{26}{\milli\meter} for Detector~2. The updated detector-specific metadata are summarized in Table~\ref{tab:material_metadata}. Crystal IDs, boule positions, carrier type and mobility, trapping length, and detailed surface-metrology records remain unavailable. This incomplete traceability limits position-dependent correlations within the original boules, but it does not alter the functional observation that the contacted Detector~2 generated voltage-dependent NTL phonons.

\begin{table*}[t]
\centering
\caption{Detector, material, fabrication, and source metadata used in the present analysis. Quantities common to both crystals are repeated where useful for clarity. Items listed as not retained should be captured in future crystal-acceptance and detector-fabrication records.}
\label{tab:material_metadata}
\begin{tabular}{lll}
\toprule
\parbox[t]{0.20\textwidth}{Quantity} &
\parbox[t]{0.34\textwidth}{Detector 1} &
\parbox[t]{0.34\textwidth}{Detector 2} \\
\midrule
\parbox[t]{0.20\textwidth}{Crystal source} &
\parbox[t]{0.34\textwidth}{USD-grown Ge} &
\parbox[t]{0.34\textwidth}{USD-grown Ge} \\[1mm]
\parbox[t]{0.20\textwidth}{Dimensions} &
\parbox[t]{0.34\textwidth}{Nominal envelope: \SI{75}{\milli\meter} diameter; \SI{10}{\milli\meter} thick; detailed as-built profile not retained} &
\parbox[t]{0.34\textwidth}{Nominal envelope: \SI{75}{\milli\meter} diameter; \SI{10}{\milli\meter} thick; detailed as-built profile not retained} \\[1mm]
\parbox[t]{0.20\textwidth}{Mass} &
\parbox[t]{0.34\textwidth}{\SI{188}{\gram}} &
\parbox[t]{0.34\textwidth}{\SI{230}{\gram}} \\[1mm]
\parbox[t]{0.20\textwidth}{Crystallographic orientation} &
\parbox[t]{0.34\textwidth}{(100)} &
\parbox[t]{0.34\textwidth}{(100)} \\[1mm]
\parbox[t]{0.20\textwidth}{Net active impurity concentration} &
\parbox[t]{0.34\textwidth}{Approximately $10^{10}~\mathrm{cm}^{-3}$ from campaign Hall measurements; carrier type and mobility were not retained} &
\parbox[t]{0.34\textwidth}{Approximately $10^{10}~\mathrm{cm}^{-3}$ from campaign Hall measurements; carrier type and mobility were not retained} \\[1mm]
\parbox[t]{0.20\textwidth}{Etch-pit density} &
\parbox[t]{0.34\textwidth}{$300<\rho_{\mathrm{EPD}}<7000~\mathrm{cm}^{-2}$; quantitative trapping length not retained} &
\parbox[t]{0.34\textwidth}{$300<\rho_{\mathrm{EPD}}<7000~\mathrm{cm}^{-2}$; quantitative trapping length not retained} \\[1mm]
\parbox[t]{0.20\textwidth}{Approximate bath temperature} &
\parbox[t]{0.34\textwidth}{\SI{10}{\milli\kelvin}} &
\parbox[t]{0.34\textwidth}{\SI{10}{\milli\kelvin}} \\[1mm]
\parbox[t]{0.20\textwidth}{Bias boundary condition} &
\parbox[t]{0.34\textwidth}{Contact-free electrode} &
\parbox[t]{0.34\textwidth}{Contact-free electrode initially; direct Al contact for the final NTL run} \\[1mm]
\parbox[t]{0.20\textwidth}{Direct Al bias film} &
\parbox[t]{0.34\textwidth}{Not used} &
\parbox[t]{0.34\textwidth}{\SI{600}{\nano\meter} nominal thickness, deposited with a modified Perkin Elmer 4400 sputter coater; deposition rate, pressure, coverage metrology, and contact resistance were not supplied} \\[1mm]
\parbox[t]{0.20\textwidth}{$^{241}$Am source} &
\parbox[t]{0.34\textwidth}{$1~\mu\mathrm{Ci}$; source-to-detector distance \SI{35.5}{\milli\meter}} &
\parbox[t]{0.34\textwidth}{$1~\mu\mathrm{Ci}$; source-to-detector distance \SI{26}{\milli\meter}} \\[1mm]
\parbox[t]{0.20\textwidth}{Bias points discussed here} &
\parbox[t]{0.34\textwidth}{Contact-free: \SIlist{0;10;20}{\volt}} &
\parbox[t]{0.34\textwidth}{Contact-free: \SIlist{0;5;10;15}{\volt}; contacted: \SIlist{0;5;10;15;20;30;40}{\volt}} \\[1mm]
\parbox[t]{0.20\textwidth}{Crystal ID and boule position} &
\parbox[t]{0.34\textwidth}{\notretained} &
\parbox[t]{0.34\textwidth}{\notretained} \\[1mm]
\parbox[t]{0.20\textwidth}{Detailed surface metrology} &
\parbox[t]{0.34\textwidth}{Visible curvature reported on the uninstrumented face; detailed polishing, chemical treatment, roughness, and flatness records not retained} &
\parbox[t]{0.34\textwidth}{Visible curvature reported on the uninstrumented face; detailed polishing, chemical treatment, roughness, and flatness records not retained} \\
\bottomrule
\end{tabular}
\end{table*}

\subsection{Phonon sensors and cryogenic readout}

\begin{figure*}[t]
\centering
\begin{minipage}[t]{0.23\textwidth}
\centering
\includegraphics[width=\linewidth]{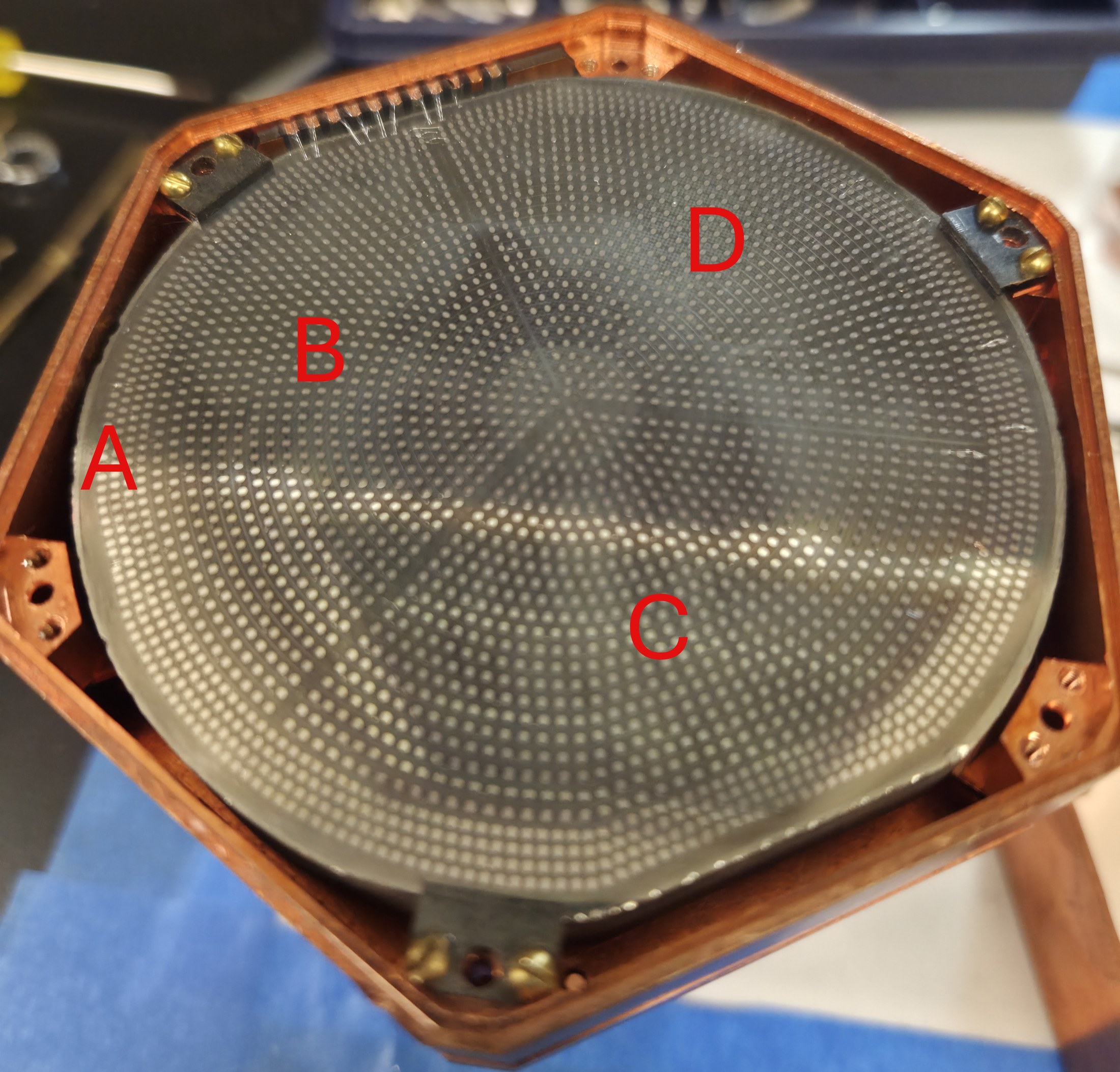}\\[-1mm]
{\small (a) Four-channel phonon face}
\end{minipage}\hfill
\begin{minipage}[t]{0.23\textwidth}
\centering
\includegraphics[width=\linewidth]{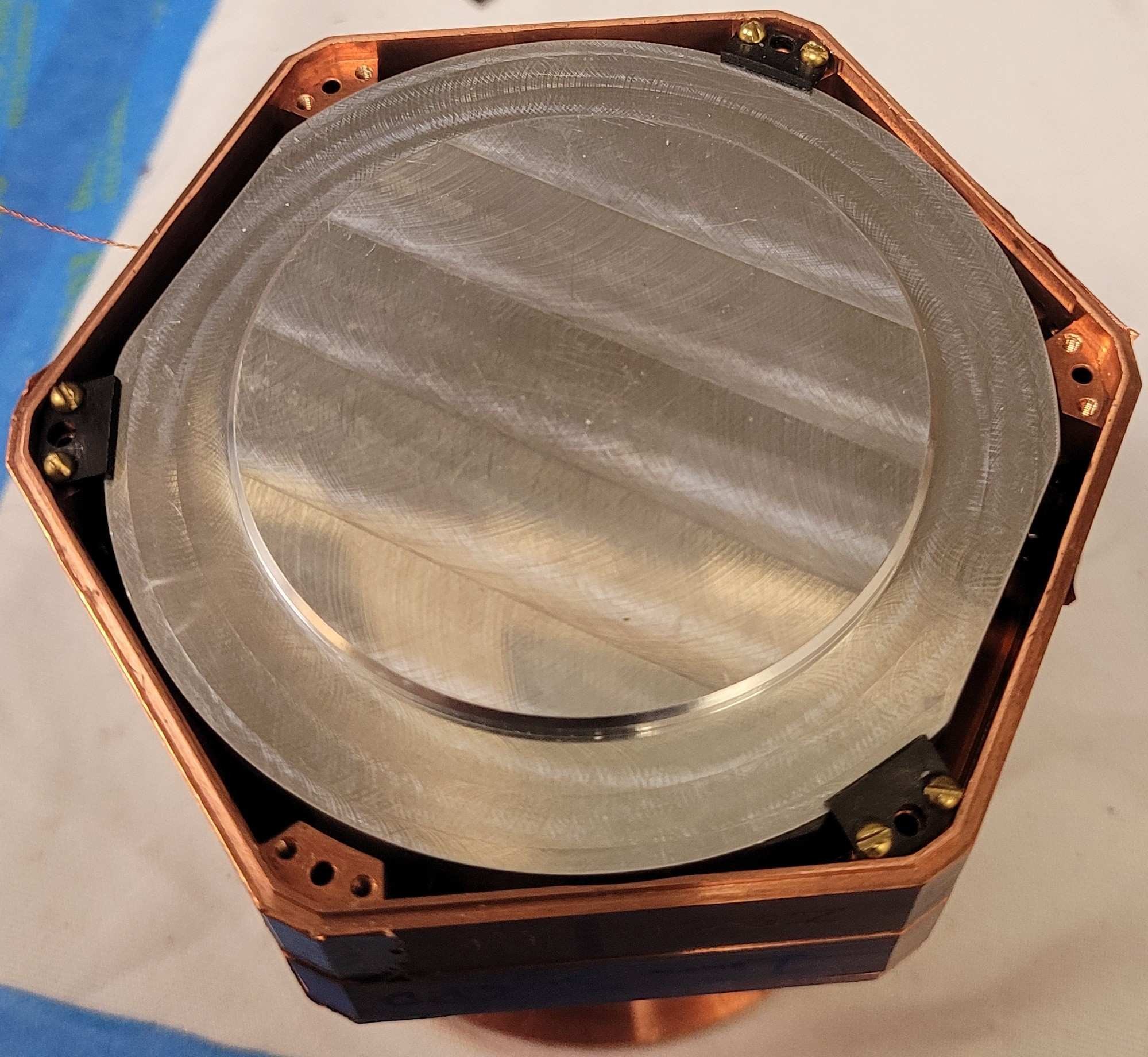}\\[-1mm]
{\small (b) Contact-free electrode}
\end{minipage}\hfill
\begin{minipage}[t]{0.23\textwidth}
\centering
\includegraphics[width=\linewidth]{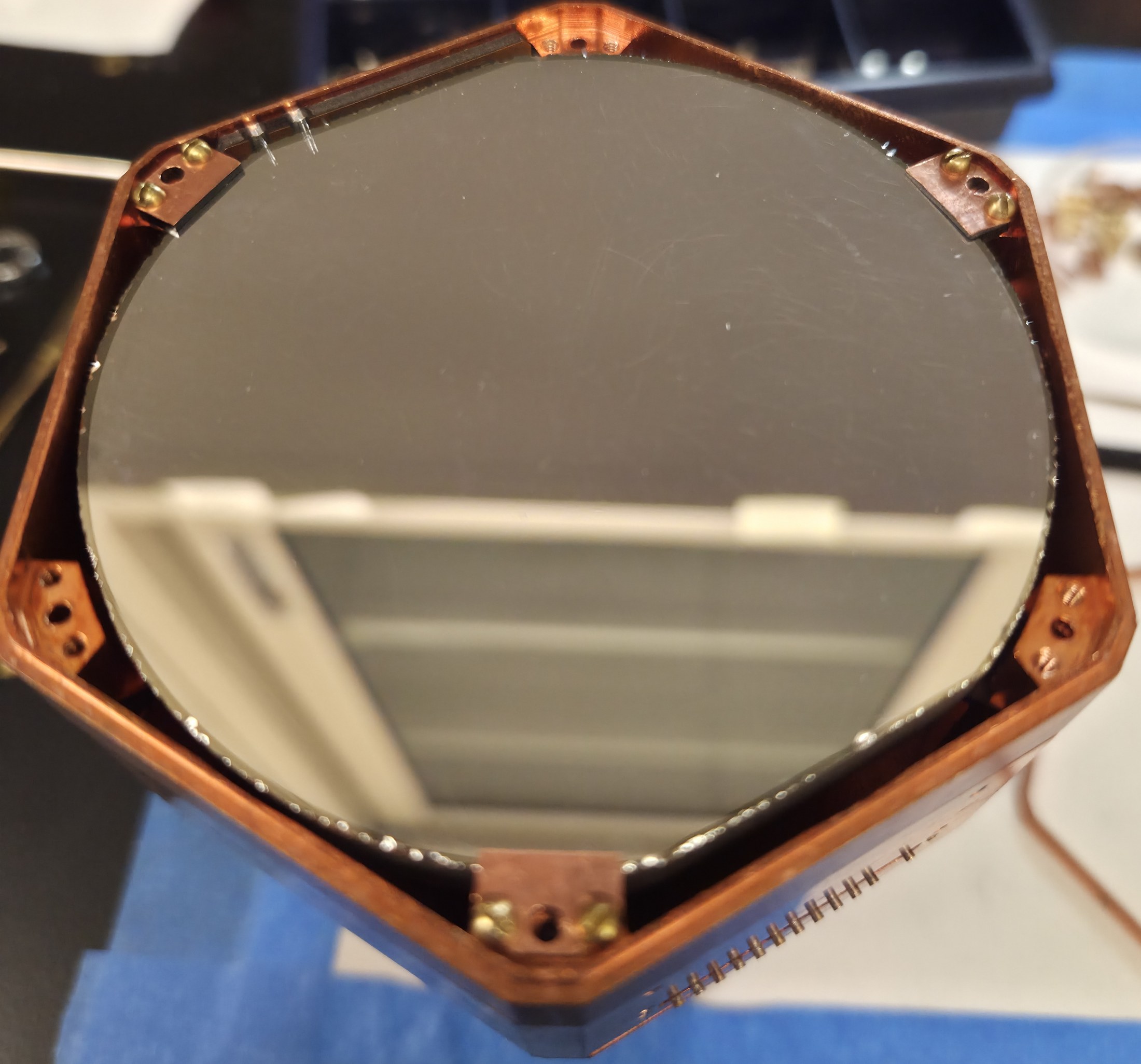}\\[-1mm]
{\small (c) Deposited Al electrode}
\end{minipage}\hfill
\begin{minipage}[t]{0.23\textwidth}
\centering
\includegraphics[width=\linewidth]{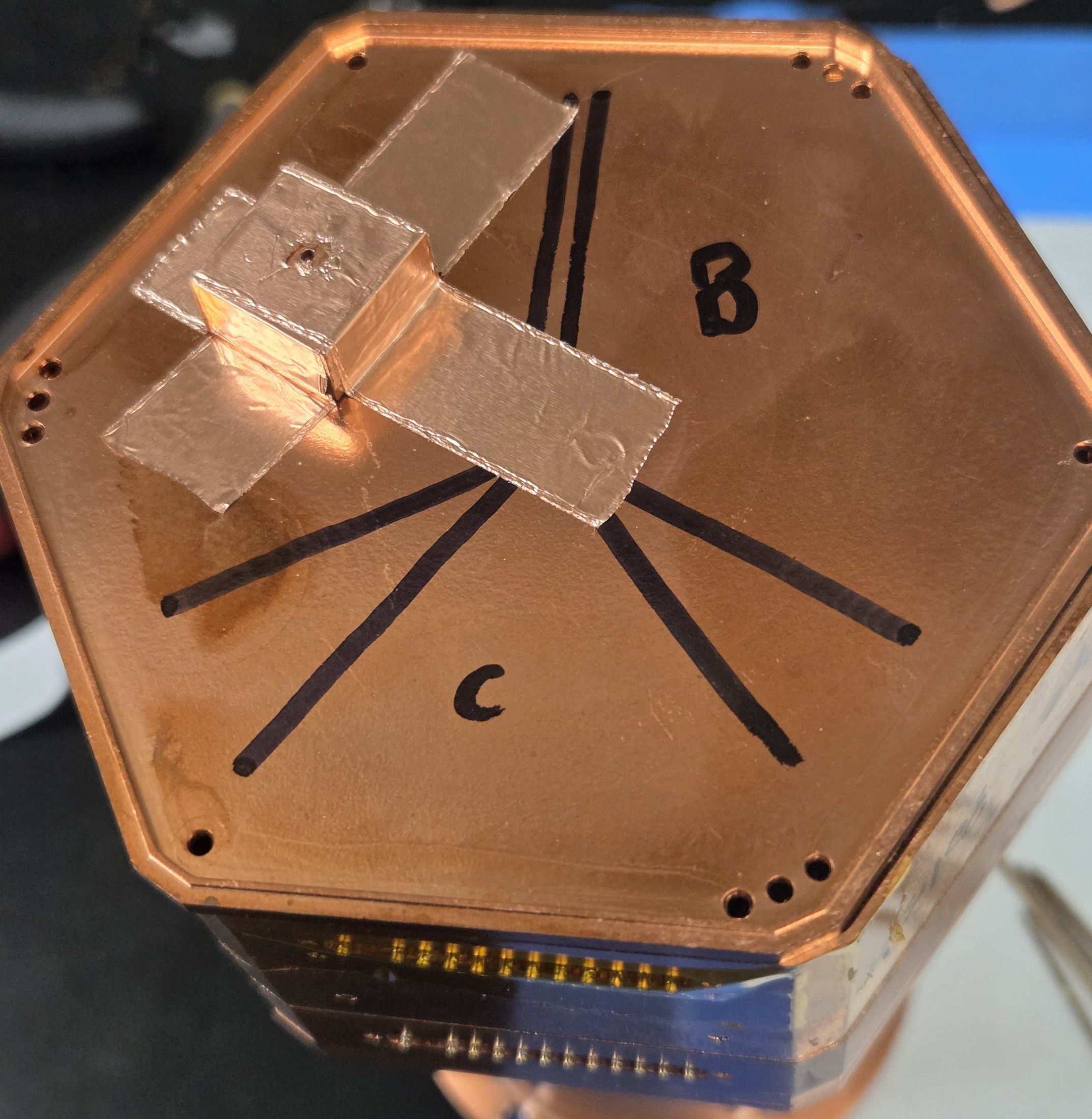}\\[-1mm]
{\small (d) $^{241}$Am source near B}
\end{minipage}
\caption{Detector configurations used in the test campaign. Detector~2 was \SI{75}{\milli\meter} in diameter, \SI{10}{\milli\meter} thick, and had a mass of \SI{230}{\gram}. It carried four phonon channels on the instrumented face. Initial operation used a mechanically separated contact-free electrode. The successful final run used a continuous \SI{600}{\nano\meter} Al electrode deposited on the opposite face with a modified Perkin Elmer 4400 sputter coater. The $1~\mu\mathrm{Ci}$ $^{241}$Am source was positioned above channel B at a source-to-detector distance of \SI{26}{\milli\meter}. Photographs are not to a common scale.}
\label{fig:detector_geometry}
\end{figure*}

Each detector was instrumented on one face with four labeled phonon channels, A--D. The sensors use superconducting transition-edge elements coupled to aluminum phonon-collection structures and are read out with SQUID electronics, following the QET/TES approach developed for cryogenic particle detection \cite{Irwin1995ETF,Irwin1995QET,Ren2021}. Figure~\ref{fig:detector_geometry} shows Detector~2, the initial contact-free electrode~\cite{Mirabolfathi2017}, the final aluminum electrode, and the source position used in the successful NTL run. The photographs are not shown to a common scale. Both substrates had nominal envelope dimensions of \SI{75}{\milli\meter} diameter and \SI{10}{\milli\meter} thickness and were operated at an approximate bath temperature of \SI{10}{\milli\kelvin}.

The parasitic or residual series resistances $R_p$ and normal resistances
$R_n$ were obtained from current-bias/current-sensor ($I_b$--$I_s$)
characterization~\cite{Anderson2011TES}, with the corresponding
resistance uncertainties listed in Table~\ref{tab:tes_parameters}. The TES transition temperatures $T_c$ were determined independently from thermal sweeps of the dilution refrigerator. For each measurement, the detector channels were first placed in the superconducting state while the mixing-chamber thermometer was read continuously; the mixing chamber was then warmed until the channels transitioned to the normal state. The procedure was repeated during cooldown as the channels transitioned from the normal state back to the superconducting state. The reported $T_c$ for each channel is the average of the warming and cooling transition temperatures. A separate uncertainty on $T_c$ is not available from this procedure and is therefore not quoted. The first detector was characterized with all four channels operational. For the second detector, channels A, B, and D were characterized initially, while channel C and revised resistance values for channel B were obtained following additional debugging and subsequent characterization.

\begin{table*}[t]
\centering
\caption{TES/QET characterization parameters for the two detector assemblies. The reported $T_c$ values are averages of the warming and cooling transition measurements; separate $T_c$ uncertainties are not available. $R_p$ and $R_n$, including their quoted uncertainties, are in m$\Omega$.}
\label{tab:tes_parameters}
\begin{tabular}{ccccc}
Detector & Channel & $T_c$ (mK) & $R_p$ (m$\Omega$) & $R_n$ (m$\Omega$) \\
\hline
1 & A & 87  & $254\pm0.337$ & $313\pm0.596$ \\
1 & B & 79  & $211\pm0.285$ & $293\pm0.515$ \\
1 & C & 104 & $46\pm0.0817$ & $302\pm0.244$ \\
1 & D & 102 & $23\pm0.0544$ & $258\pm0.187$ \\
\hline
2 & A & 75 & $45\pm0.0335$ & $253\pm0.271$ \\
2 & B & 61 & $20\pm0.0180$ & $311\pm0.0353$ \\
2 & C & 78 & $21\pm0.0182$ & $378\pm0.0392$ \\
2 & D & 66 & $37\pm0.0293$ & $214\pm0.232$ \\
\end{tabular}
\end{table*}

\subsection{Contact-free bias tests}

Detector~1 was first operated in February 2025 with a contact-free electrode placed below the face opposite the phonon sensors. A $1~\mu\mathrm{Ci}$ $^{241}$Am source was used at a source-to-detector distance of \SI{35.5}{\milli\meter}. Several months of measurements were performed. The June 2025 run was the only period in which all four channels operated simultaneously; other runs were limited by shorts, broken SQUID or QET lines, or unstable channel behavior. Spectra acquired at \SI{0}{\volt}, \SI{10}{\volt}, and \SI{20}{\volt} did not show a reproducible voltage-dependent peak shift.

\begin{figure}[t]
\centering
\includegraphics[width=\columnwidth]{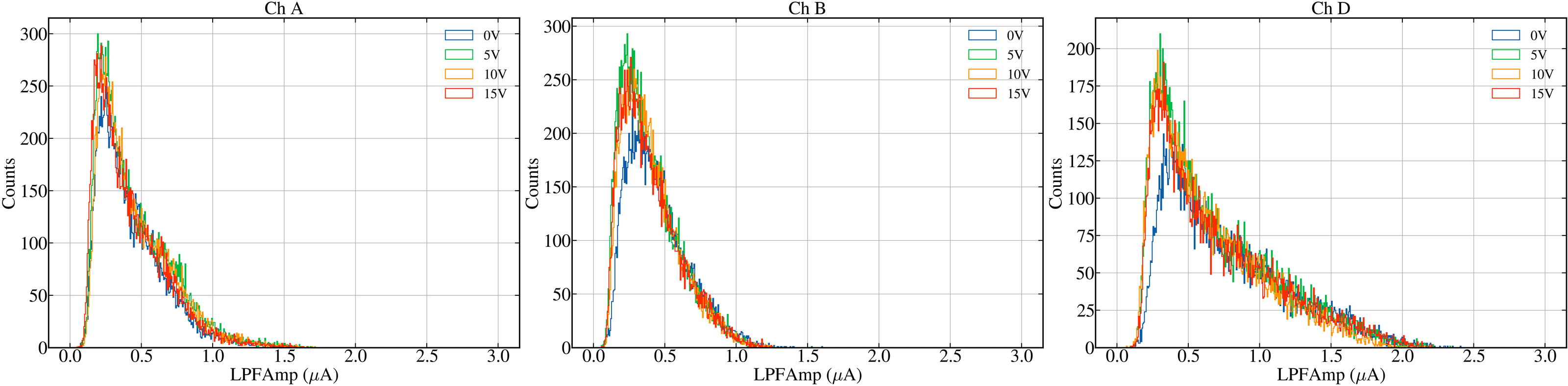}
\caption{Amplitude spectra from the second detector in the contact-free configuration at \SI{0}{\volt}, \SI{5}{\volt}, \SI{10}{\volt}, and \SI{15}{\volt}. No systematic NTL-induced peak displacement was resolved. The absent fourth panel denotes \emph{channel C unavailable} during this run and must not be interpreted as a zero-response measurement.}
\label{fig:contactfree}
\end{figure}

Detector~2 was tested with the same contact-free geometry. The $1~\mu\mathrm{Ci}$ $^{241}$Am source was located \SI{26}{\milli\meter} from the detector. Spectra acquired at \SI{0}{\volt}, \SI{5}{\volt}, \SI{10}{\volt}, and \SI{15}{\volt} likewise showed no measurable NTL amplification, as shown in Figure~\ref{fig:contactfree}. LEDs were subsequently installed for detector neutralization, but the absence of gain persisted. Three mechanisms were considered: incomplete charge transport caused by trapping, leakage or surface charge that screened the internal field, and an NTL signal obscured by environmental and readout noise. A pronounced baseline response when \SI{20}{\volt} was applied indicated that the voltage reached the detector assembly, but the campaign did not independently measure the leakage current or the time-dependent internal field.

\begin{figure*}[t]
\centering
\begin{minipage}[t]{0.48\textwidth}
\centering
\includegraphics[width=\linewidth]{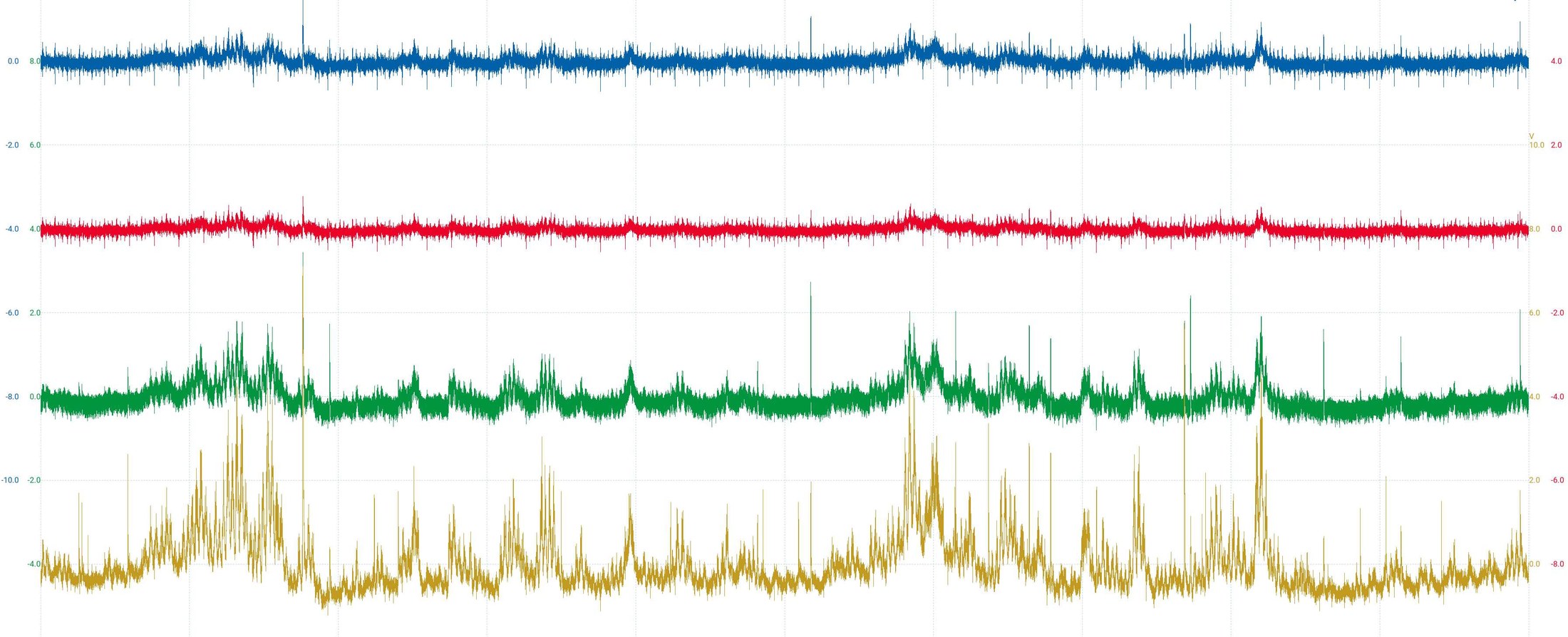}\\[-1mm]
{\small (a) Compressor on}
\end{minipage}\hfill
\begin{minipage}[t]{0.48\textwidth}
\centering
\includegraphics[width=\linewidth]{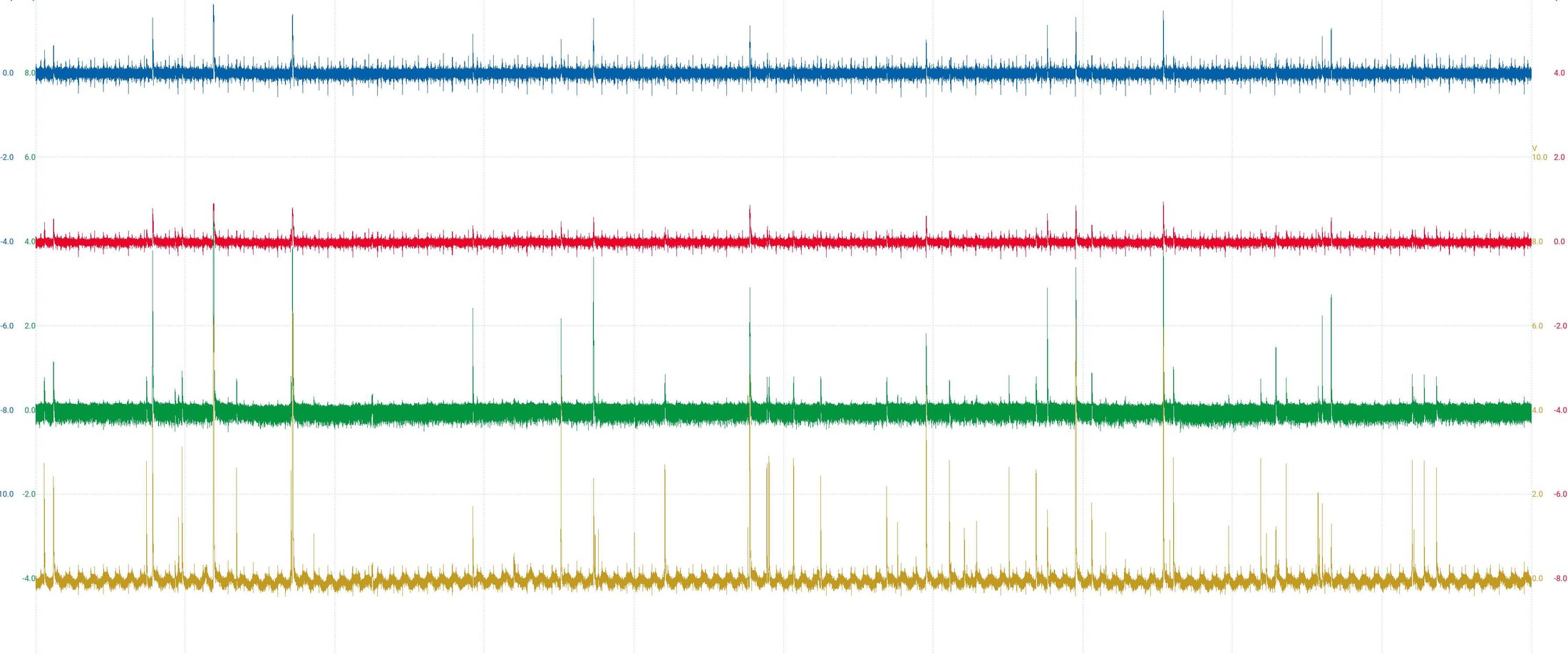}\\[-1mm]
{\small (b) Compressor off}
\end{minipage}
\caption{One-second waveform segments illustrating the environmental-noise limitation. Compressor operation generated strong microphonic structure; with the compressor off, the baseline became substantially quieter, although narrow-band electrical pickup remained in at least one channel.}
\label{fig:microphonics}
\end{figure*}

The dilution refrigerator introduced substantial compressor-driven microphonic noise. As shown in Figure~\ref{fig:microphonics}, turning the compressor off strongly reduced the broad vibrational structure, but sustained operation in that state was not possible because the refrigerator warmed. A separate \SI{120}{\hertz} pickup, consistent with ground-loop contamination, appeared in some channels. These observations show that failure to resolve gain in the contact-free runs cannot be attributed uniquely to the bulk crystal.

\subsection{Contacted-electrode final run}

Because curvature on the uninstrumented crystal face made a patterned electrode difficult to install, a continuous aluminum film was deposited directly on that face for the final June 2026 run. The film was deposited with a modified Perkin Elmer 4400 sputter coater to a nominal thickness of \SI{600}{\nano\meter}. The deposition rate, sputter pressure, film-uniformity or coverage metrology, and contact resistance were not supplied with the present record. The deposited film replaced an uncertain capacitive/contact-free boundary condition with direct electrical contact, thereby improving field establishment while also changing phonon reflection and absorption at the electrode surface. Detector~2 was operated at an approximate bath temperature of \SI{10}{\milli\kelvin}. The $1~\mu\mathrm{Ci}$ $^{241}$Am source was first placed near channel D and was then moved above channel B; the source-to-detector distance was \SI{26}{\milli\meter}. Data were acquired at detector biases of \SIlist{0;5;10;15;20;30;40}{\volt}. The contacted configuration therefore tested whether the earlier null result was caused by an intrinsic inability of the crystal to transport charge or by detector fabrication, bias boundary conditions, and readout limitations.

All SQUID and QET lines were repaired before the run, but the available cryogenic time was limited to two days. Channel A left its superconducting transition when \SI{5}{\volt} was applied, plausibly because of local heating. Channel D could not be stabilized in its transition region on the available timescale. Channels B and C remained usable. Channel B remained usable throughout the 0--40~V scan, but the archived records do not preserve a point-by-point determination of its TES operating resistance $R_0$ as a function of detector bias. The loss of two channels prevents a detector-wide performance claim, but it does not invalidate the narrower channel-B test of whether the USD-grown crystal could support charge drift and NTL phonon generation.

\subsection{Data acquisition, event selection, and peak extraction}
\label{sec:event_selection}

Waveforms were digitized at \SI{1}{\mega\hertz} with a \SI{4}{\milli\second} record length. For the final contacted run, the total-phonon channel, denoted PT, was formed as the sum of the two usable phonon channels, B and C, and was used for triggering and event finding. The PT waveform was convolved with a representative pulse template, and candidate pulses were identified as peaks in the convolution whose prominence exceeded the analysis threshold. A region around each candidate peak was then selected for reconstruction. Regions containing multiple pulses were rejected to suppress pileup.

The event-selection sequence contained three additional classes of cuts. First, events for which the reduced-quantity (RQ) extraction returned NaN or infinite values, or otherwise satisfied conditions indicating an unreliable reconstruction, were removed as basic quality failures. Second, the optimal-filter pulse time was required to fall within a $100~\mu\mathrm{s}$ window centered on the trigger, i.e., within $50~\mu\mathrm{s}$ on either side. Third, reconstruction-quality cuts removed events with large sums of squared error (SSE) or large $\chi^2$. The SSE was computed between the low-pass-filtered pulse and the pulse template after normalizing both to unit maximum amplitude; this cut primarily rejects noise transients and can also remove pileup. The $\chi^2$ selection provides additional rejection of pileup and saturated events.

The pre-cut trigger counts and the accepted-event counts after the selections above are reported in Table~\ref{tab:peak_fits}. The pre-cut counts vary substantially with bias, from 146,540 at \SI{0}{\volt} to 450,769 at \SI{5}{\volt}, even though the acquisition times are similar. The trigger algorithm was unchanged across the bias settings, and its prominence threshold lies within the noise region, producing a large population of pre-cut triggers. A voltage-dependent trigger-efficiency curve was not calculated. Consequently, neither the pre-cut trigger counts nor the accepted-event counts are interpreted as measurements of the $^{241}$Am interaction rate or as quantitative trigger/selection efficiencies, and they are not used as evidence for NTL amplification.

For accepted events, the optimal filter was used to determine pulse timing only. That timing estimate provided an approximate location for the low-pass-filter (LPF) amplitude-extraction algorithm to search for the pulse maximum. The LPF amplitude, rather than the optimal-filter amplitude, was used as the quantitative pulse-amplitude estimator. This choice was made because the optimal-filter amplitude estimator is most appropriate for approximately Gaussian, stationary noise, whereas these data contain substantial structured and time-dependent noise contributions, including compressor-driven microphonics and \SI{120}{\hertz} pickup. The dominant channel-B peak in the LPF-amplitude histogram at each bias was then fit with a Gaussian,
\begin{equation}
g(x)=A\exp\left[-\frac{1}{2}\left(\frac{x-\mu}{\sigma}\right)^2\right],
\label{eq:gaussian_peak}
\end{equation}
over the bias-dependent intervals listed in Table~\ref{tab:peak_fits}. These are the same LPF-amplitude distributions and Gaussian curves displayed in Figure~\ref{fig:ntl_spectra}. The restricted intervals confine the fit to the vicinity of the dominant peak, and the present analysis uses a single Gaussian without an explicit background term. Histogram-bin uncertainties were not used as fit weights. Accordingly, a weighted-fit $\chi^2/\mathrm{ndf}$ is not reported. The quoted uncertainty on $\mu$ is the statistical centroid uncertainty returned by the unweighted Gaussian fit; it does not include fit-model or background systematics. Fit residuals and a comparison with a background-inclusive model are not used in the present analysis. A dedicated robustness study in which the Gaussian-only fits are repeated with altered fit windows or an explicit local background term is also not available from the archived analysis products. Consequently, no numerical fit-model systematic is assigned to the centroids. The qualitative NTL conclusion relies on the large and monotonic overall displacement of the dominant channel-B peak across the bias scan, whereas the precise centroid-to-energy conversion retains the fit-model and calibration limitations stated here.

\begin{table*}[t]
\centering
\caption{Detector~2 contacted-run live time, pre-cut trigger count, number of accepted pulses after event selection, Gaussian fit interval, and fitted channel-B low-pass-filter (LPF) peak position. Fit intervals and peak positions are expressed in $\mu$A for readability; the underlying supplied values were in amperes. The Gaussian fits were unweighted, and the quoted $\delta\mu_{\mathrm{stat}}$ is the fit-returned statistical centroid uncertainty. A weighted-fit $\chi^2/\mathrm{ndf}$ and background-model systematic are therefore not quoted.}
\label{tab:peak_fits}
\small
\begin{tabular}{cccccc}
\toprule
Bias (V) & Live time (s) & Pre-cut triggers & Accepted pulses & Fit interval ($\mu$A) & $\mu\pm\delta\mu_{\mathrm{stat}}$ ($\mu$A) \\
\midrule
0  & 1995 & 146\,540 & 30\,798 & 2.208--2.900   & $2.377\pm0.00618$ \\
5  & 1995 & 450\,769 & 57\,169 & 4.705--5.385   & $4.895\pm0.00711$ \\
10 & 2128 & 187\,814 & 68\,217 & 7.150--8.067   & $7.381\pm0.01147$ \\
15 & 1995 & 383\,822 & 68\,474 & 9.866--11.220  & $10.150\pm0.02388$ \\
20 & 1995 & 371\,120 & 74\,737 & 11.920--13.130 & $12.140\pm0.02476$ \\
30 & 1995 & 400\,007 & 71\,880 & 14.170--16.030 & $14.640\pm0.02376$ \\
40 & 1995 & 319\,718 & 72\,300 & 15.730--18.190 & $16.420\pm0.02910$ \\
\bottomrule
\end{tabular}
\end{table*}

The fitted centroid moves from $(2.377\pm0.006)~\mu\mathrm{A}$ at \SI{0}{\volt} to $(16.420\pm0.029)~\mu\mathrm{A}$ at \SI{40}{\volt}. The statistical uncertainty on an individual peak position is only approximately 0.15--0.26\% of the corresponding centroid. Consequently, peak-position statistical error bars are smaller than the plotted symbols at the scale used for the bias-dependence figures. These small statistical errors should not, however, be interpreted as the full uncertainty on reconstructed absorbed energy or phonon-collection efficiency.

\subsection{Data reduction and uncertainty treatment}

For a voltage-biased TES operated under electrothermal feedback, the energy absorbed by the sensor can be estimated from the current pulse \cite{Irwin1995ETF,Ren2021}. The analysis used
\begin{equation}
\begin{split}
E_{\mathrm{abs}}={}&\left(1-\frac{2R_L}{R_L+R_0}\right)
I_B R_{\mathrm{sh}}\int \delta I(t)\,dt \\
&+R_L\int \delta I^2(t)\,dt,
\end{split}
\label{eq:eabs}
\end{equation}
where $R_L=R_{\mathrm{sh}}+R_p$, $R_{\mathrm{sh}}=\SI{20}{\milli\ohm}$, $R_0$ is the TES operating resistance, $I_B=84.4~\mu\mathrm{A}$ for channel B, and $\delta I(t)$ is the reconstructed pulse current. The archived reduction used the nominal operating-point approximation $R_0\simeq0.5R_n$ for channel B. A bias-by-bias determination of $R_0$ is not available, so any detector-bias dependence of the TES operating point is not propagated into the reconstructed absorbed-energy uncertainty. Equation~\eqref{eq:eabs} converts the current-pulse integral into the energy absorbed by the TES. The apparent total phonon collection efficiency was defined as
\begin{equation}
\eta_{\mathrm{tot}}(V)=\frac{E_{\mathrm{abs}}(V)}{E_{\mathrm{ph}}(V)}.
\label{eq:phce}
\end{equation}
Equation~\eqref{eq:phce}, evaluated with the ideal phonon energy from Equation~\eqref{eq:ntl_gain}, was used to determine the voltage-dependent apparent collection efficiency. The dominant $^{241}$Am calibration line was taken as $E_R=\SI{59.54}{\kilo\electronvolt}$ \cite{IAEAAm241}.

The reanalysis provides statistical uncertainties on the fitted LPF peak positions, as reported in Table~\ref{tab:peak_fits}. These uncertainties quantify how precisely the Gaussian centroid is determined after the stated event selections. For the $E_{\mathrm{abs}}$ calculation, the pulse-current integrals in Eq.~\eqref{eq:eabs} were not evaluated directly from the raw pulses. Instead, the generated pulse template was scaled by the fitted peak centroid $\mu$, and the scaled template was integrated. This procedure removes residual baseline offsets from the energy reconstruction: although the integral of an ideal baseline is zero, a nonzero baseline in an individual raw waveform could otherwise bias the result. Consequently, baseline resolution is not included as a separate term in Eq.~\eqref{eq:eabs}. A complete point-by-point uncertainty on $E_{\mathrm{abs}}$ and $\eta_{\mathrm{tot}}$ nevertheless requires propagation of the statistical uncertainty in $\mu$, the template and absolute amplitude/energy calibration uncertainties and their drift, repeatability across acquisitions, possible detector-bias dependence of the TES operating point, and the relevant uncertainties and covariances among $R_p$, $R_n$, $R_0$, $I_B$, and other TES-response parameters. Accordingly, the reconstructed absorbed energies and collection efficiencies reported below should be interpreted as nominal values under the stated TES operating-point assumption.

Because the required pulse-integral, calibration, and detector-response uncertainties have not yet been propagated, the regression in Sec.~\ref{sec:efficiency_fit} remains unweighted. The statistical LPF peak-centroid errors are small enough to be visually unresolved at the scale of the bias-dependence plots, but using them alone as weights for the derived-energy fit would substantially understate the total uncertainty. Quoted regression standard errors therefore describe the scatter of the tabulated energy points about the linear model and are not substitutes for a complete calibration and systematic-error budget.

\section{Observation of NTL amplification}
\label{sec:results}

\subsection{Bias-dependent spectra}

\begin{figure*}[t]
\centering
\includegraphics[width=0.99\textwidth]{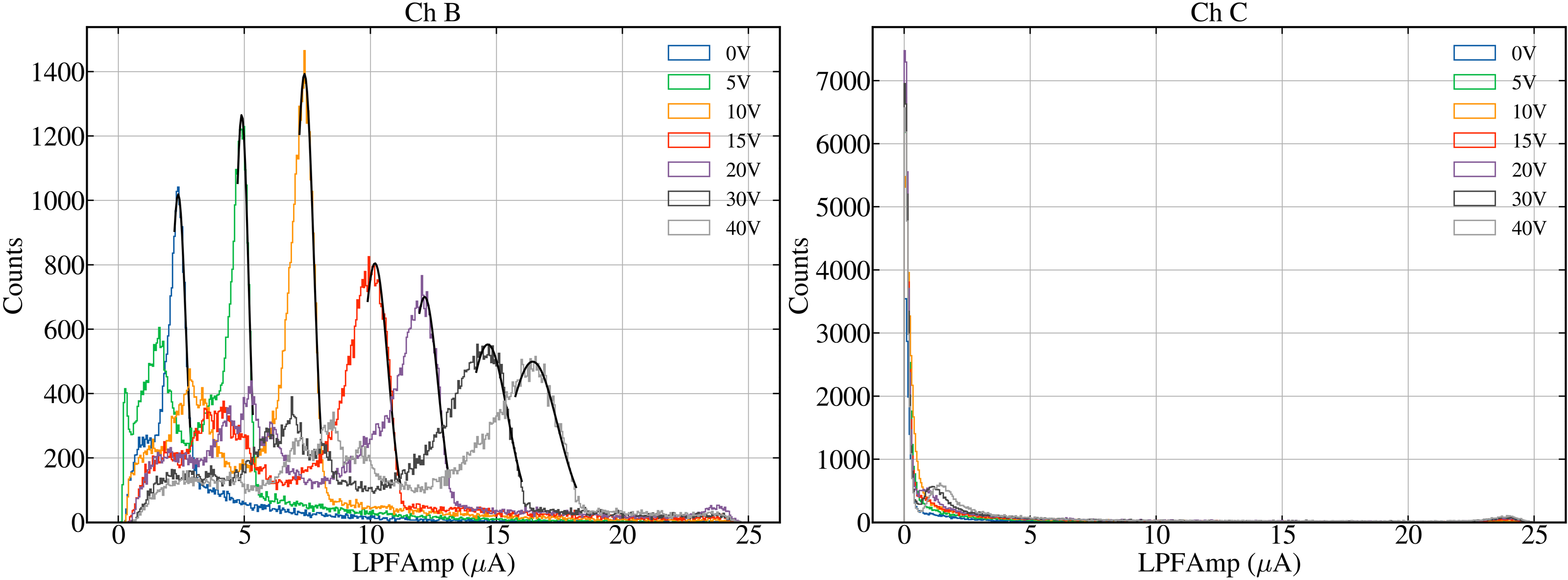}
\caption{Final-run low-pass-filter (LPF) amplitude spectra for channels B (left) and C (right) at \SIlist{0;5;10;15;20;30;40}{\volt}. The curves overlaid on the channel-B histograms are the Gaussian fits used to obtain the quantitative peak positions reported in Table~\ref{tab:peak_fits}. The optimal filter was used only to estimate pulse timing and thereby provide an approximate search location for LPF amplitude extraction. Channel B shows clear voltage-dependent NTL amplification. The weak response in channel C is consistent with the $^{241}$Am source placement over channel B and local phonon collection.}
\label{fig:ntl_spectra}
\end{figure*}

Figure~\ref{fig:ntl_spectra} shows the final LPF-amplitude spectra from channels B and C and the Gaussian curves used in the quantitative peak extraction. The channel-B centroid values and statistical uncertainties listed in Table~\ref{tab:peak_fits} are obtained from these LPF-amplitude distributions; the optimal filter was used only for pulse timing, as described in Sec.~\ref{sec:event_selection}. Channel B exhibits a sequence of well-separated spectral maxima that move to larger pulse amplitude as the bias increases from \SI{0}{\volt} to \SI{40}{\volt}. The Gaussian centroid increases from $(2.377\pm0.006)~\mu\mathrm{A}$ at \SI{0}{\volt} to $(16.420\pm0.029)~\mu\mathrm{A}$ at \SI{40}{\volt}. The voltage-dependent displacement is much larger than the statistical centroid uncertainties at every bias setting, while the absolute calibration and detector-response systematics remain to be quantified. In contrast, channel C shows little gain. This channel asymmetry is consistent with the source position and local phonon partition: events produced predominantly beneath channel B deliver a larger fraction of their athermal phonon energy to that channel. The observation in channel B is sufficient for the narrow functional question addressed here because NTL production requires mobile charge carriers to traverse a nonzero potential in the USD-grown germanium.

\subsection{TES absorbed energy and apparent phonon collection}

\begin{figure}[t]
\centering
\includegraphics[width=\columnwidth]{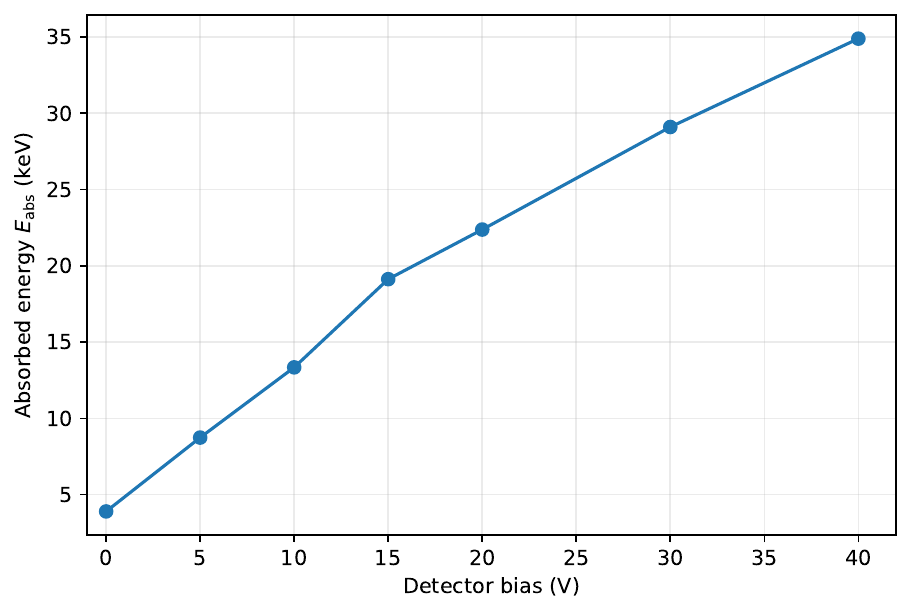}
\caption{Nominal reconstructed channel-B absorbed energy as a function of detector bias. The monotonic overall increase provides a direct visualization of NTL amplification. Statistical uncertainties on the fitted low-pass-filter peak centroids are below 0.3\% (Table~\ref{tab:peak_fits}) and would be smaller than the plotting symbols, but they are not propagated as $E_{\mathrm{abs}}$ error bars because the uncertainties of the pulse-current integrals, calibration, detector-bias dependence of the TES operating point, and other TES-response parameters required by Eq.~\eqref{eq:eabs} have not yet been fully propagated.}
\label{fig:eabs_bias}
\end{figure}

\begin{figure}[t]
\centering
\includegraphics[width=\columnwidth]{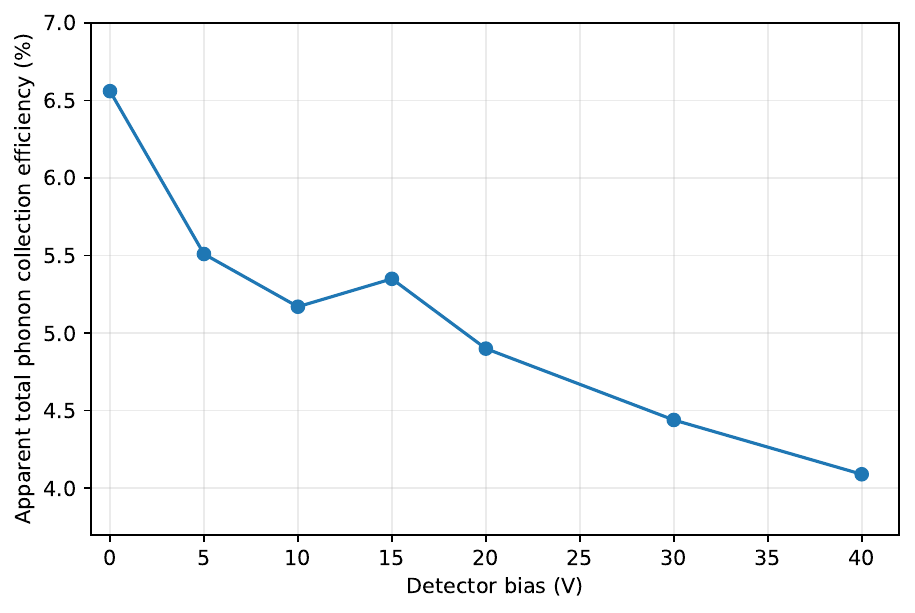}
\caption{Apparent total phonon collection efficiency obtained by dividing the measured TES absorbed energy by the ideal electron-recoil phonon energy in Eq.~\eqref{eq:ntl_gain}. The overall decreasing trend with bias, with a modest upward fluctuation at \SI{15}{\volt}, indicates that prompt and NTL phonons are not collected with identical effective efficiency in this geometry. The statistical LPF peak-centroid uncertainties are smaller than the plotting symbols, but a defensible error bar on $\eta_{\mathrm{tot}}$ also requires propagation of the pulse-current-integral, energy-calibration, and TES-response uncertainties; these are therefore not shown as pointwise error bars.}
\label{fig:phce_bias}
\end{figure}

Table~\ref{tab:ntl_results} gives the nominal reconstructed channel-B absorbed energies and apparent collection efficiencies calculated using Eqs.~\eqref{eq:ntl_gain}--\eqref{eq:phce}. Under the nominal TES operating-point assumption, the absorbed energy rises with bias, reaching approximately \SI{34.9}{\kilo\electronvolt} at \SI{40}{\volt}. Relative to the \SI{0}{\volt} response, the nominal reconstructed peak energy is enhanced by a factor of approximately 8.9 at \SI{40}{\volt}. The ideal electron-recoil NTL energy increases more rapidly than the measured channel-B energy, so the apparent total collection efficiency decreases from \SI{6.56}{\percent} to \SI{4.09}{\percent} over the same range. Figure~\ref{fig:eabs_bias} displays the measured increase of $E_{\mathrm{abs}}$ with bias, while Figure~\ref{fig:phce_bias} shows the corresponding overall decrease of $\eta_{\mathrm{tot}}$ obtained from Equation~\eqref{eq:phce}, with a modest upward fluctuation at \SI{15}{\volt}.

\begin{table}[t]
\caption{Channel-B NTL response reconstructed from the final June 2026 run. The listed $E_{\mathrm{abs}}$ and $\eta_{\mathrm{tot}}$ values are nominal results obtained with the archived TES-response parameters and the assumption $R_0\simeq0.5R_n$. Statistical uncertainties on the fitted low-pass-filter peak positions are reported separately in Table~\ref{tab:peak_fits}. A complete uncertainty on $E_{\mathrm{abs}}$ and $\eta_{\mathrm{tot}}$ requires propagation of the pulse-current-integral, calibration, TES operating-point, and other TES-response uncertainties that are not yet quantified.}
\label{tab:ntl_results}
\begin{tabular}{ccc}
Bias (V) & $E_{\mathrm{abs}}$ (keV) & $\eta_{\mathrm{tot}}$ (\%) \\
\hline
0  & 3.90  & 6.56 \\
5  & 8.74  & 5.51 \\
10 & 13.34 & 5.17 \\
15 & 19.12 & 5.35 \\
20 & 22.37 & 4.90 \\
30 & 29.09 & 4.44 \\
40 & 34.88 & 4.09 \\
\end{tabular}
\end{table}

\subsection{Effective prompt and NTL phonon response coefficients}
\label{sec:efficiency_fit}

\begin{figure}[t]
\centering
\includegraphics[width=\columnwidth]{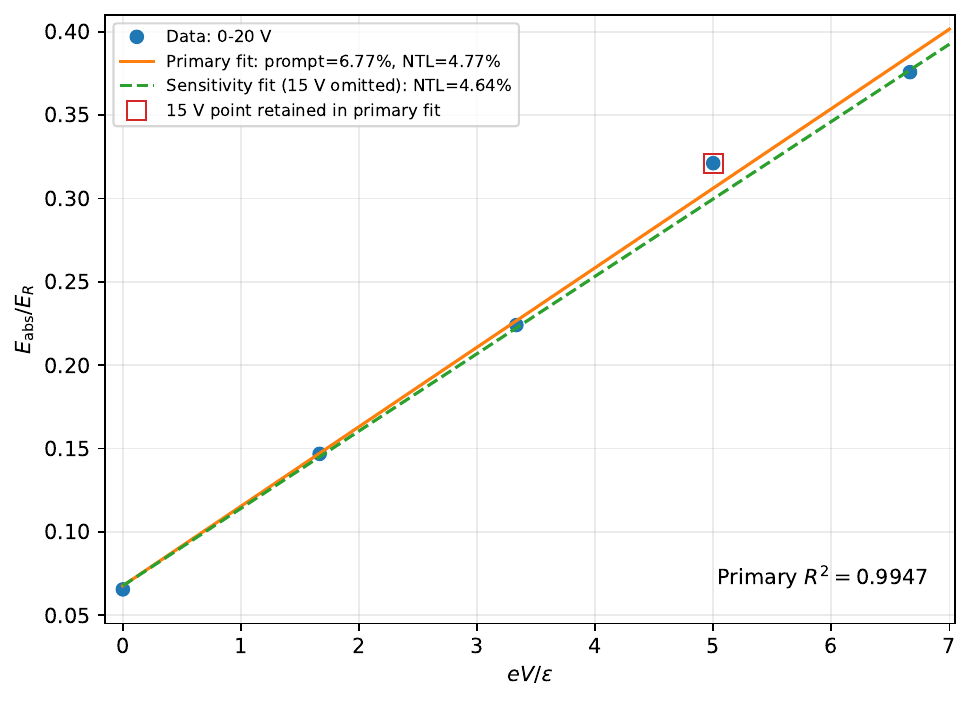}
\caption{Low-bias decomposition of the channel-B response using Eq.~\eqref{eq:two_eff}. The primary fit includes all five points from \SI{0}{\volt} through \SI{20}{\volt}; the dashed sensitivity fit reproduces the original selection that omitted the \SI{15}{\volt} point. The \SI{30}{\volt} and \SI{40}{\volt} measurements are excluded from this coefficient-extraction figure because Eq.~\eqref{eq:two_eff} is treated as a low-bias constant-efficiency parameterization; the all-seven-point diagnostic is reported in Table~\ref{tab:fit_comparison}. The intercept and slope are interpreted as effective channel-B prompt-phonon and NTL-phonon response coefficients, respectively. Statistical uncertainties on the fitted pulse-amplitude centroids are smaller than the symbols, but the fit is intentionally unweighted because those centroid errors do not include the calibration and TES-response uncertainties of the derived absorbed energies.}
\label{fig:eff_fit}
\end{figure}

A phenomenological decomposition defines effective channel-B collection coefficients for prompt recoil phonons and field-generated NTL phonons:
\begin{equation}
E_{\mathrm{abs}}=\eta_{\mathrm{prompt}}E_R+
\eta_{\mathrm{NTL}}E_R\frac{eV}{\varepsilon}.
\label{eq:two_eff}
\end{equation}
Equation~\eqref{eq:two_eff} implies that a plot of $E_{\mathrm{abs}}/E_R$ against $eV/\varepsilon$ has intercept $\eta_{\mathrm{prompt}}$ and slope $\eta_{\mathrm{NTL}}$. In this manuscript Eq.~\eqref{eq:two_eff} is used as a low-bias phenomenological decomposition with approximately voltage-independent effective coefficients. The primary coefficient extraction is therefore restricted to the five points from \SI{0}{\volt} through \SI{20}{\volt}; the \SI{30}{\volt} and \SI{40}{\volt} points remain part of the spectra and $\eta_{\mathrm{tot}}$ analyses but are not used to define the low-bias coefficients because the high-bias data show increasing departure from a single constant-efficiency linear description. This is a model-domain choice rather than pointwise statistical outlier rejection. Within the selected low-bias interval, all five points are retained with no a priori point rejection. The unweighted fit gives
\begin{align}
\eta_{\mathrm{prompt}} &= (6.77\pm0.82)\%,
\label{eq:eta_prompt}\\
\eta_{\mathrm{NTL}} &= (4.77\pm0.20)\%.
\label{eq:eta_ntl}
\end{align}
The uncertainties in Eqs.~\eqref{eq:eta_prompt} and \eqref{eq:eta_ntl} are regression standard errors only. The primary fit has $R^2=0.9947$. To quantify sensitivity to the elevated \SI{15}{\volt} point, a second fit reproduces the original four-point selection (\SIlist{0;5;10;20}{\volt}) and gives $\eta_{\mathrm{prompt}}=(6.77\pm0.19)\%$, $\eta_{\mathrm{NTL}}=(4.64\pm0.05)\%$, and $R^2=0.9998$. As a separate model-domain diagnostic, an unweighted fit of the same linear form to all seven points from \SI{0}{\volt} through \SI{40}{\volt} gives $\eta_{\mathrm{prompt}}=(9.19\pm1.59)\%$, $\eta_{\mathrm{NTL}}=(3.91\pm0.22)\%$, and $R^2=0.9842$. The shift in both effective coefficients and the reduced linearity over the full range support treating Eq.~\eqref{eq:two_eff} as a low-bias parameterization rather than a global detector model. Table~\ref{tab:fit_comparison} compares these analyses, and Figure~\ref{fig:eff_fit} displays the two low-bias fits. The four-point result is treated as a sensitivity check rather than the primary result because the available data do not support a statistical outlier-rejection criterion.

\begin{table}[t]
\caption{Unweighted linear-fit diagnostics for Eq.~\eqref{eq:two_eff}. The primary and four-point sensitivity fits probe the low-bias regime; the all-seven-point row is included only to test the validity of extending the same constant-coefficient model through \SI{40}{\volt}. Quoted uncertainties are regression standard errors and do not include calibration, detector-response, or TES-parameter systematics.}
\label{tab:fit_comparison}
\begin{tabular}{lccc}
Fit selection & $\eta_{\mathrm{prompt}}$ (\%) & $\eta_{\mathrm{NTL}}$ (\%) & $R^2$ \\
\hline
Primary: 0--20 V & $6.77\pm0.82$ & $4.77\pm0.20$ & 0.9947 \\
Sensitivity: 0, 5, 10, 20 V & $6.77\pm0.19$ & $4.64\pm0.05$ & 0.9998 \\
All seven: 0--40 V & $9.19\pm1.59$ & $3.91\pm0.22$ & 0.9842 \\
\end{tabular}
\end{table}

\section{Discussion}
\label{sec:discussion}

\subsection{Crystal quality and detector fabrication are inseparable in an NTL test}

The principal experimental conclusion is that the tested USD-grown germanium supports field-driven charge transport and NTL phonon generation at cryogenic temperature. The evidence is the systematic displacement of the channel-B spectral maximum with increasing bias, the increase of reconstructed absorbed energy shown in Figure~\ref{fig:eabs_bias}, and the approximately linear low-bias dependence described by Eqs.~\eqref{eq:eta_prompt} and \eqref{eq:eta_ntl}. NTL phonons cannot be generated without mobile charge carriers traversing an internal potential. The successful contacted-electrode run therefore strongly disfavors the strongest bulk-material failure hypothesis: that the tested crystal is intrinsically incapable of useful charge transport in this NTL detector.

Equally important, the campaign demonstrates why crystal quality cannot be inferred from an NTL null result without first validating the fabricated device. Observable gain requires a chain of conditions: the bulk must permit carrier motion; surfaces and electrodes must suppress injection and uncontrolled leakage; the applied voltage must establish the intended internal field; charge must traverse a significant fraction of that potential; generated phonons must propagate to the instrumented face; QET/TES channels must remain in transition; and the cryogenic environment must be quiet enough to reconstruct the pulses. A defect at any one link can mask otherwise adequate crystal quality. This system dependence is consistent with previous high-voltage-detector studies and with the strong influence of contact processing and geometry observed in conventional USD-grown HPGe detectors \cite{Ponce2020,Albakry2022,Wilson2024,Wei2018,Meng2019}.

The history of the campaign provides a comparative qualitative demonstration. Two detector assemblies showed no measurable gain in the contact-free geometry, while the second crystal showed clear gain after direct aluminum contact was added and the sensor/readout problems were repaired. Because the bulk crystal was unchanged, the transition from a null result to NTL amplification argues against an intrinsic inability of that crystal to transport charge and implicates detector-level boundary conditions, field establishment, fabrication quality, and readout performance as important variables. Several changes occurred between the unsuccessful and successful configurations, including the electrode boundary condition, readout repair, and source placement, so the present campaign does not isolate a unique causal change. The result also does not prove that trapping is negligible; it shows that any residual trapping was not sufficient to prevent useful charge drift through the tested potential.

\subsection{Proof-of-principle compatibility with SuperCDMS-style HV detectors}

The Hall-derived impurity concentration of approximately $10^{10}~\mathrm{cm}^{-3}$, (100) orientation, and measured etch-pit-density range $300<\rho_{\mathrm{EPD}}<7000~\mathrm{cm}^{-2}$ establish that the tested substrates possess material characteristics relevant to detector fabrication. A Hall measurement or impurity-density specification alone, however, is not sufficient to assess a crystal for a SuperCDMS-style HV detector. The material must survive mechanical processing and microfabrication, accept stable electrode and QET structures, operate at the approximately \SI{10}{\milli\kelvin} bath temperature used here, and support voltage-dependent charge drift without an immediate collapse of the measured response. In the present campaign, a SuperCDMS-style detector fabricated from a USD-grown crystal produced clear NTL amplification through \SI{40}{\volt}. The result therefore demonstrates proof-of-principle compatibility of the tested material with this detector class and supports continued fabrication studies using USD-grown crystals.

This statement is intentionally narrower than production acceptance of a complete SuperCDMS detector. Production qualification would additionally require detector-wide field uniformity, all-channel yield, calibrated leakage current, repeatable neutralization, controlled charge trapping and impact ionization, stable long-duration operation, quantified energy resolution, and reproducible performance over multiple crystals. The present work demonstrates a foundational material--device compatibility milestone and identifies the fabrication controls required to advance toward detector-level acceptance criteria.

\subsection{Interpretation of the lower effective NTL response coefficient}

The primary fit gives an effective NTL-phonon response coefficient approximately 30\% lower than the prompt-phonon coefficient. Several detector-level effects can produce this difference without implying poor bulk crystal quality. Prompt recoil phonons and NTL phonons are generated with different spatial and temporal distributions. The continuous aluminum electrode changes the boundary condition on the uninstrumented face and can absorb phonons that would otherwise reflect toward the QET side. The source was localized above channel B, making the response sensitive to position-dependent field lines and channel partition. Surface trapping, field nonuniformity, or incomplete carrier traversal can also reduce the Luke energy actually generated. These effects are recognized in high-voltage phonon detectors, where trapping, impact ionization, and low-energy backgrounds require dedicated modeling and calibration \cite{Ponce2020,Albakry2022,Wilson2024,Neog2022}.

The overall decrease of $\eta_{\mathrm{tot}}$ with voltage shown in Figure~\ref{fig:phce_bias}, including the modest upward fluctuation at \SI{15}{\volt}, should therefore not be interpreted as a voltage-dependent deterioration of intrinsic crystal quality. It is a detector-level observable that reflects the combined effects of bulk charge transport, electrode geometry, phonon propagation, aluminum absorption, QET coverage, TES operating conditions and possible saturation, as well as pulse-reconstruction performance. A future comparison of multiple USD crystals should hold these detector variables fixed and include a position-dependent field-and-phonon model.

\subsection{Limitations and next validation steps}

The measurement is a proof of principle rather than a complete SuperCDMS production qualification. The final high-quality data are dominated by one inner phonon channel. The run did not provide a direct leakage-current measurement, bias-polarity reversal, a calibrated position scan, a quantitative charge-trapping length, a long-term stability study, or a bias-by-bias measurement of the channel-B TES operating resistance. Statistical uncertainties on the Gaussian LPF peak positions are quantified in Table~\ref{tab:peak_fits}, but the uncertainties of the pulse-current integrals used in Eq.~\eqref{eq:eabs}, the absolute calibration uncertainty and drift, and the propagated TES energy-reconstruction systematics are not yet available. The Gaussian peak fits were performed without histogram-bin uncertainty weights, so a weighted-fit $\chi^2/\mathrm{ndf}$ is not reported; fit residuals, a comparison with a background-inclusive fit model, and a quantitative fit-window/background robustness study are also not included in the present analysis. The large bias-dependent displacement of the dominant channel-B maximum is visually and numerically clear, but no separate numerical systematic is assigned for possible centroid shifts under alternative peak models. The pre-cut trigger counts are now reported in Table~\ref{tab:peak_fits}, but a voltage-dependent trigger-efficiency curve was not calculated, so the bias dependence of trigger and accepted-event counts cannot be converted into an interaction-rate or selection-efficiency measurement. Two channels were unavailable in the final run. Detector nominal dimensions, reported masses, operating temperature, source activity and geometry, live time, sampling, pulse-record length, and the Detector~2 aluminum deposition method and thickness are now documented; the detailed as-built geometry required to reconcile Detector~1's nominal envelope dimensions with its reported mass is not retained. Figure~\ref{fig:ntl_spectra} shows the LPF-amplitude histograms and Gaussian fits used for the quantitative Table~\ref{tab:peak_fits} peak extraction; the optimal filter supplies timing information only. Crystal IDs, boule positions, carrier type and mobility, detailed surface processing, Al deposition rate/pressure and contact resistance, and a complete detector-response uncertainty budget remain incomplete. These limitations prevent claims of detector-wide phonon-collection uniformity, quantitative carrier lifetime, or direct equivalence to the best HVeV energy resolutions and collection efficiencies \cite{Agnese2018HVeV,Ren2021}.

The most direct next experiment is a controlled A/B comparison using planarized crystal faces and a reproducible patterned electrode. The protocol should include: (i) crystal IDs and boule-position traceability; (ii) room-temperature and cryogenic Hall measurements linked to each substrate; (iii) dislocation and surface-flatness metrology; (iv) documented electrode thickness, coverage, deposition conditions, and contact resistance; (v) continuous leakage-current and bias-current monitoring; (vi) LED neutralization before each bias step; (vii) both bias polarities and repeated voltage cycles; (viii) compressor vibration isolation and removal of the \SI{120}{\hertz} ground loop; (ix) stable operation of all four QET channels with continuous TES operating-point monitoring; (x) a collimated source scan across the detector; (xi) repeat peak extraction with controlled fit-window variations and background-inclusive models; and (xii) a field-and-phonon Monte Carlo constrained by the measured channel partition. Repeating this protocol on several USD-grown crystals would convert the present proof of principle into a statistically defensible crystal-acceptance study.

\section{Conclusion}

A SuperCDMS-style cryogenic phonon detector fabricated from USD-grown germanium exhibited clear NTL amplification after a continuous \SI{600}{\nano\meter} aluminum bias electrode was deposited on the face opposite the QET sensors with a modified Perkin Elmer 4400 sputter coater. Channel B showed a systematic voltage-dependent response from \SI{0}{\volt} to \SI{40}{\volt}, with the fitted low-pass-filter peak moving from $(2.377\pm0.006)~\mu\mathrm{A}$ to $(16.420\pm0.029)~\mu\mathrm{A}$. Under the nominal TES operating-point assumption used in the archived reduction, the reconstructed absorbed energy increased by approximately a factor of 8.9. The primary low-bias fit, which retains all points from \SI{0}{\volt} through \SI{20}{\volt}, yielded effective channel-B prompt- and NTL-phonon response coefficients of $(6.77\pm0.82)\%$ and $(4.77\pm0.20)\%$, respectively. The peak-position uncertainties are statistical fit errors; the quoted response-coefficient uncertainties quantify regression scatter only and do not include the full calibration, TES operating-point, or energy-reconstruction systematics. A full measurement uncertainty still requires propagation of the pulse-current-integral, calibration, and TES-response uncertainties.

The central result is broader than the observation of a moving spectral peak. Both tested substrates had nominal envelope dimensions of \SI{75}{\milli\meter} diameter and \SI{10}{\milli\meter} thickness; their reported masses were \SI{188}{\gram} and \SI{230}{\gram}. The tested material had a Hall-derived net impurity concentration of approximately $10^{10}~\mathrm{cm}^{-3}$, a (100) orientation, and an etch-pit density in the range $300<\rho_{\mathrm{EPD}}<7000~\mathrm{cm}^{-2}$. The successful Detector~2 run demonstrates that a crystal produced by the USD university-based growth program can be mechanically processed, integrated with SuperCDMS-style phonon sensors and a high-voltage electrode, cooled to an approximate bath temperature of \SI{10}{\milli\kelvin}, and made to sustain the charge transport required for NTL phonon generation. The tested crystal therefore demonstrates proof-of-principle material--device compatibility for continued SuperCDMS-style HV detector development. It is not yet a production-qualified detector or a general acceptance test for all USD-grown crystals.

The campaign also supports a necessary qualification principle: NTL performance is determined jointly by crystal growth and detector fabrication. The initial contact-free null results did not demonstrate inadequate crystal quality; the successful contacted run showed that electrode geometry, surface and leakage control, QET/TES yield, phonon collection, and environmental noise can dominate whether material capability is observed. Future work should repeat a controlled and fully traceable fabrication process on multiple USD-grown crystals with full channel yield, leakage monitoring, field mapping, source scans, bias-polarity reversal, uncertainty propagation, and long-term stability measurements. The present result closes the first proof-of-principle loop from USD crystal growth to measurable NTL amplification and provides a technically justified basis for the next stage of the USD crystal-growth and detector-development program.

\acknowledgments
The authors acknowledge the technical contributions of members of the University of South Dakota germanium crystal-growth and detector-development program and the Mirabolfathi group at Texas A\&M University. This work was supported in part by the U.S. Department of Energy under Award No.~DE-SC0024519 and by the U.S. National Science Foundation under Award No.~OIA-2437416 and Award No.~OISE-1743790. The authors also acknowledge the cryogenic, SQUID-readout, detector-integration, and thin-film fabrication support that enabled the measurements.

\section*{Author contributions}
D.M. conceived and coordinated the USD crystal-growth and detector-validation effort, interpreted the results, and prepared the manuscript. G.G. performed the cryogenic detector testing and primary data analysis. N.M. supervised the cryogenic detector integration and measurements, contributed to interpretation, and reviewed the manuscript. W. B. and M.P. contributed to detector integration and thin-film fabrication, provided fabrication metadata for the contacted-electrode configuration, and reviewed the manuscript. All authors approved the final manuscript.

\section*{Data availability}
The processed numerical values supporting the principal conclusions are reported in Tables~\ref{tab:peak_fits}, \ref{tab:ntl_results}, and \ref{tab:fit_comparison}. Table~\ref{tab:peak_fits} reports the Detector~2 live time, pre-cut trigger count, accepted-event count, Gaussian fit interval, and statistical LPF peak-position uncertainty for every contacted-run bias setting. Table~\ref{tab:material_metadata} reports the detector dimensions and masses, approximate \SI{10}{\milli\kelvin} operating temperature, source activity and source-to-detector distances, and the Detector~2 Al deposition method and nominal thickness. Raw waveform, calibration, and analysis-level data are not publicly deposited at present. Reasonable requests should be directed to the corresponding author and will be handled in accordance with applicable collaboration data-sharing requirements. The numerical values needed to reproduce the regressions reported here are provided in Tables~\ref{tab:ntl_results} and \ref{tab:fit_comparison}. A complete uncertainty propagation for the derived absorbed energies and collection efficiencies requires uncertainties on the pulse-current integrals used in Eq.~\eqref{eq:eabs}, absolute calibration uncertainty and drift, and TES-response covariance information, which are not quantified in the present manuscript.

\section*{Competing interests}
The authors declare no competing interests.

\bibliographystyle{unsrtnat}
\bibliography{references}

\begin{thebibliography}{32}
\providecommand{\natexlab}[1]{#1}
\providecommand{\url}[1]{\texttt{#1}}
\expandafter\ifx\csname urlstyle\endcsname\relax
  \providecommand{\doi}[1]{doi: #1}\else
  \providecommand{\doi}{doi: \begingroup \urlstyle{rm}\Url}\fi

\bibitem[Neganov and Trofimov(1985)]{Neganov1985}
B.~S. Neganov and V.~N. Trofimov.
\newblock Colorimetric method measuring ionizing radiation.
\newblock Otkrytiya i Izobreteniya, 1985.
\newblock USSR patent no. 1037771.

\bibitem[Luke(1988)]{Luke1988}
P.~N. Luke.
\newblock Voltage-assisted calorimetric ionization detector.
\newblock \emph{Journal of Applied Physics}, 64\penalty0 (12):\penalty0 6858--6860, 1988.
\newblock \doi{10.1063/1.341976}.

\bibitem[{SuperCDMS Collaboration}(2014)]{Agnese2014}
{SuperCDMS Collaboration}.
\newblock Search for low-mass weakly interacting massive particles using voltage-assisted calorimetric ionization detection in the {SuperCDMS} experiment.
\newblock \emph{Physical Review Letters}, 112:\penalty0 041302, 2014.
\newblock \doi{10.1103/PhysRevLett.112.041302}.

\bibitem[{SuperCDMS Collaboration}(2016)]{Agnese2016}
{SuperCDMS Collaboration}.
\newblock New results from the search for low-mass weakly interacting massive particles with the {CDMS} low ionization threshold experiment.
\newblock \emph{Physical Review Letters}, 116:\penalty0 071301, 2016.
\newblock \doi{10.1103/PhysRevLett.116.071301}.

\bibitem[{SuperCDMS Collaboration}(2018{\natexlab{a}})]{Agnese2018CDMSlite}
{SuperCDMS Collaboration}.
\newblock Low-mass dark matter search with {CDMSlite}.
\newblock \emph{Physical Review D}, 97:\penalty0 022002, 2018{\natexlab{a}}.
\newblock \doi{10.1103/PhysRevD.97.022002}.

\bibitem[{SuperCDMS Collaboration}(2018{\natexlab{b}})]{Agnese2018HVeV}
{SuperCDMS Collaboration}.
\newblock First dark matter constraints from a {SuperCDMS} single-charge sensitive detector.
\newblock \emph{Physical Review Letters}, 121:\penalty0 051301, 2018{\natexlab{b}}.
\newblock \doi{10.1103/PhysRevLett.121.051301}.
\newblock Erratum: Phys. Rev. Lett. 122, 069901 (2019).

\bibitem[{EDELWEISS Collaboration}(2020)]{Arnaud2020}
{EDELWEISS Collaboration}.
\newblock First germanium-based constraints on {Sub-MeV} dark matter with the {EDELWEISS} experiment.
\newblock \emph{Physical Review Letters}, 125:\penalty0 141301, 2020.
\newblock \doi{10.1103/PhysRevLett.125.141301}.

\bibitem[{SuperCDMS Collaboration}(2017)]{Agnese2017}
{SuperCDMS Collaboration}.
\newblock Projected sensitivity of the {SuperCDMS} {SNOLAB} experiment.
\newblock \emph{Physical Review D}, 95:\penalty0 082002, 2017.
\newblock \doi{10.1103/PhysRevD.95.082002}.

\bibitem[Freedman(1974)]{Freedman1974}
Daniel~Z. Freedman.
\newblock Coherent effects of a weak neutral current.
\newblock \emph{Physical Review D}, 9\penalty0 (5):\penalty0 1389--1392, 1974.
\newblock \doi{10.1103/PhysRevD.9.1389}.

\bibitem[{COHERENT Collaboration}(2017)]{Akimov2017}
{COHERENT Collaboration}.
\newblock Observation of coherent elastic neutrino-nucleus scattering.
\newblock \emph{Science}, 357\penalty0 (6356):\penalty0 1123--1126, 2017.
\newblock \doi{10.1126/science.aao0990}.

\bibitem[Colaresi et~al.(2022)Colaresi, Collar, Hossbach, Lewis, and Yocum]{Colaresi2022}
J.~Colaresi, J.~I. Collar, T.~W. Hossbach, C.~M. Lewis, and K.~M. Yocum.
\newblock Measurement of coherent elastic neutrino-nucleus scattering from reactor antineutrinos.
\newblock \emph{Physical Review Letters}, 129:\penalty0 211802, 2022.
\newblock \doi{10.1103/PhysRevLett.129.211802}.

\bibitem[Ackermann et~al.(2025)Ackermann, Bonet, Bonhomme, Buck, F{\"u}lber, Hakenm{\"u}ller, Hempfling, Heusser, Lindner, Maneschg, Ni, Rank, Rink, S{\'a}nchez~Garc{\'i}a, Stalder, Strecker, Wink, and Woenckhaus]{Ackermann2025}
N.~Ackermann, H.~Bonet, A.~Bonhomme, C.~Buck, K.~F{\"u}lber, J.~Hakenm{\"u}ller, J.~Hempfling, G.~Heusser, M.~Lindner, W.~Maneschg, K.~Ni, M.~Rank, T.~Rink, E.~S{\'a}nchez~Garc{\'i}a, I.~Stalder, H.~Strecker, R.~Wink, and J.~Woenckhaus.
\newblock Direct observation of coherent elastic antineutrino--nucleus scattering.
\newblock \emph{Nature}, 643:\penalty0 1229--1233, 2025.
\newblock \doi{10.1038/s41586-025-09322-2}.

\bibitem[Neog et~al.(2022)Neog, Mahapatra, Mirabolfathi, Platt, Jastram, Agnolet, Kubik, Chen, and Mohanty]{Neog2022}
H.~Neog, R.~Mahapatra, N.~Mirabolfathi, M.~Platt, A.~Jastram, G.~Agnolet, A.~Kubik, H.~Chen, and B.~Mohanty.
\newblock Phonon-mediated high-voltage detector with background rejection for low-mass dark matter and reactor coherent neutrino scattering experiments.
\newblock \emph{Nuclear Instruments and Methods in Physics Research Section A}, 1033:\penalty0 166707, 2022.
\newblock \doi{10.1016/j.nima.2022.166707}.

\bibitem[Ponce et~al.(2020)Ponce, Page, Brink, Cabrera, Cherry, Fink, Kurinsky, Partridge, Pyle, Sadoulet, Serfass, Stanford, Watkins, Yellin, and Young]{Ponce2020}
F.~Ponce, W.~Page, P.~L. Brink, B.~Cabrera, M.~Cherry, C.~Fink, N.~Kurinsky, R.~Partridge, M.~Pyle, B.~Sadoulet, B.~Serfass, C.~Stanford, S.~L. Watkins, S.~Yellin, and B.~A. Young.
\newblock Measuring the impact ionization and charge trapping probabilities in {SuperCDMS} {HVeV} phonon sensing detectors.
\newblock \emph{Physical Review D}, 101:\penalty0 031101, 2020.
\newblock \doi{10.1103/PhysRevD.101.031101}.

\bibitem[{SuperCDMS Collaboration}(2022)]{Albakry2022}
{SuperCDMS Collaboration}.
\newblock Investigating the sources of low-energy events in a {SuperCDMS-HVeV} detector.
\newblock \emph{Physical Review D}, 105:\penalty0 112006, 2022.
\newblock \doi{10.1103/PhysRevD.105.112006}.

\bibitem[Wilson et~al.(2024)Wilson, Zaytsev, von Krosigk, Alkhatib, Buchanan, Chen, Diamond, Figueroa-Feliciano, Harms, Hong, Kennard, Kurinsky, Mahapatra, Mirabolfathi, Novati, Platt, Ren, Sattari, Schmidt, Wang, Zatschler, Zhang, and Zuniga]{Wilson2024}
M.~J. Wilson, A.~Zaytsev, B.~von Krosigk, I.~Alkhatib, M.~Buchanan, R.~Chen, M.~D. Diamond, E.~Figueroa-Feliciano, S.~A.~S. Harms, Z.~Hong, K.~T. Kennard, N.~A. Kurinsky, R.~Mahapatra, N.~Mirabolfathi, V.~Novati, M.~Platt, R.~Ren, A.~Sattari, B.~Schmidt, Y.~Wang, S.~Zatschler, E.~Zhang, and A.~Zuniga.
\newblock Improved modeling of detector response effects in phonon-based crystal detectors used for dark matter searches.
\newblock \emph{Physical Review D}, 109:\penalty0 112018, 2024.
\newblock \doi{10.1103/PhysRevD.109.112018}.

\bibitem[Ren et~al.(2021)Ren, Bathurst, Chang, Chen, Fink, Hong, Kurinsky, Mast, Mishra, Novati, Spahn, Meyer~zu Theenhausen, Watkins, Williams, Wilson, Zaytsev, Bauer, Bunker, Figueroa-Feliciano, Hollister, Hsu, Lukens, Mahapatra, Mirabolfathi, Nebolsky, Platt, Ponce, Pyle, Reynolds, and Saab]{Ren2021}
R.~Ren, C.~Bathurst, Y.-Y. Chang, R.~Chen, C.~W. Fink, Z.~Hong, N.~A. Kurinsky, N.~Mast, N.~Mishra, V.~Novati, G.~Spahn, H.~Meyer~zu Theenhausen, S.~L. Watkins, Z.~Williams, M.~J. Wilson, A.~Zaytsev, D.~Bauer, R.~Bunker, E.~Figueroa-Feliciano, M.~Hollister, L.~Hsu, P.~Lukens, R.~Mahapatra, N.~Mirabolfathi, B.~Nebolsky, M.~Platt, F.~Ponce, M.~Pyle, T.~Reynolds, and T.~Saab.
\newblock Design and characterization of a phonon-mediated cryogenic particle detector with an {eV}-scale threshold and 100 {keV}-scale dynamic range.
\newblock \emph{Physical Review D}, 104:\penalty0 032010, 2021.
\newblock \doi{10.1103/PhysRevD.104.032010}.

\bibitem[Wei et~al.(2018)Wei, Meng, Li, Liu, Wang, Mei, Yang, Mei, and Zhang]{Wei2018}
W.-Z. Wei, X.-H. Meng, Y.-Y. Li, J.~Liu, G.-J. Wang, H.~Mei, G.~Yang, D.-M. Mei, and C.~Zhang.
\newblock Investigation of amorphous germanium contact properties with planar detectors made from {USD}-grown germanium crystals.
\newblock \emph{Journal of Instrumentation}, 13:\penalty0 P12026, 2018.
\newblock \doi{10.1088/1748-0221/13/12/P12026}.

\bibitem[Meng et~al.(2019)Meng, Wang, Wagner, Mei, Wei, Liu, Yang, and Mei]{Meng2019}
X.-H. Meng, G.-J. Wang, M.-D. Wagner, H.~Mei, W.-Z. Wei, J.~Liu, G.~Yang, and D.-M. Mei.
\newblock Fabrication and characterization of high-purity germanium detectors with amorphous germanium contacts.
\newblock \emph{Journal of Instrumentation}, 14:\penalty0 P02019, 2019.
\newblock \doi{10.1088/1748-0221/14/02/P02019}.

\bibitem[Panth et~al.(2020)Panth, Liu, Abt, Liu, Schulz, Wei, Mei, Mei, et~al.]{Panth2020}
R.~Panth, J.~Liu, I.~Abt, X.~Liu, O.~Schulz, W.-Z. Wei, H.~Mei, D.-M. Mei, et~al.
\newblock Characterization of high-purity germanium detectors with amorphous germanium contacts in cryogenic liquids.
\newblock \emph{European Physical Journal C}, 80:\penalty0 667, 2020.
\newblock \doi{10.1140/epjc/s10052-020-8235-9}.

\bibitem[Haller et~al.(1981)Haller, Hansen, and Goulding]{Haller1981}
E.~E. Haller, W.~L. Hansen, and F.~S. Goulding.
\newblock Physics of ultra-pure germanium.
\newblock \emph{Advances in Physics}, 30\penalty0 (1):\penalty0 93--138, 1981.
\newblock \doi{10.1080/00018738100101357}.

\bibitem[Abrosimov et~al.(2020)Abrosimov, Czupalla, Dropka, Fischer, Gybin, Irmscher, Janicsk{\'o}-Cs{\'a}thy, Juda, Kayser, Miller, Pietsch, and Kie{\ss}ling]{Abrosimov2020}
N.~V. Abrosimov, M.~Czupalla, N.~Dropka, J.~Fischer, A.~Gybin, K.~Irmscher, J.~Janicsk{\'o}-Cs{\'a}thy, U.~Juda, S.~Kayser, W.~Miller, M.~Pietsch, and F.~M. Kie{\ss}ling.
\newblock Technology development of high purity germanium crystals for radiation detectors.
\newblock \emph{Journal of Crystal Growth}, 532:\penalty0 125396, 2020.
\newblock \doi{10.1016/j.jcrysgro.2019.125396}.

\bibitem[Wang et~al.(2012)Wang, Sun, Yang, Xiang, Guan, Mei, Keller, and Chan]{Wang2012}
Guojian Wang, Yongchen Sun, Gang Yang, Wenzhi Xiang, Yutong Guan, Dongming Mei, C.~Keller, and Y.-D. Chan.
\newblock Development of large size high-purity germanium crystal growth.
\newblock \emph{Journal of Crystal Growth}, 352\penalty0 (1):\penalty0 27--30, 2012.
\newblock \doi{10.1016/j.jcrysgro.2012.01.018}.

\bibitem[Wang et~al.(2015)Wang, Mei, Mei, Guan, and Yang]{Wang2015USD}
Guojian Wang, Hao Mei, Dongming Mei, Yutong Guan, and Gang Yang.
\newblock High purity germanium crystal growth at the university of south dakota.
\newblock \emph{Journal of Physics: Conference Series}, 606:\penalty0 012012, 2015.
\newblock \doi{10.1088/1742-6596/606/1/012012}.

\bibitem[Bhattarai et~al.(2024)Bhattarai, Mei, Budhathoki, Dong, and Warren]{Bhattarai2024}
Sanjay Bhattarai, Dongming Mei, Narayan Budhathoki, Kunming Dong, and Austin Warren.
\newblock Investigating influential parameters for high-purity germanium crystal growth.
\newblock \emph{Crystals}, 14\penalty0 (2):\penalty0 177, 2024.
\newblock \doi{10.3390/cryst14020177}.

\bibitem[Prem et~al.(2026)Prem, Mei, Bhattarai, Budhathoki, and Chhetri]{Prem2026}
Athul Prem, Dongming Mei, Sanjay Bhattarai, Narayan Budhathoki, and Sunil Chhetri.
\newblock Machine-learning optimization of detector-grade yield in high-purity germanium crystal growth.
\newblock \emph{Journal of Crystal Growth}, 694:\penalty0 128766, 2026.
\newblock \doi{10.1016/j.jcrysgro.2026.128766}.

\bibitem[Mei(2024)]{Mei2024}
Dongming Mei.
\newblock Enhancing sensitivity in ge-based rare-event physics experiments through underground crystal growth and detector fabrication.
\newblock \emph{arXiv e-prints}, 2024.
\newblock \doi{10.48550/arXiv.2409.03580}.

\bibitem[Irwin(1995)]{Irwin1995ETF}
K.~D. Irwin.
\newblock An application of electrothermal feedback for high resolution cryogenic particle detection.
\newblock \emph{Applied Physics Letters}, 66\penalty0 (15):\penalty0 1998--2000, 1995.
\newblock \doi{10.1063/1.113674}.

\bibitem[Irwin et~al.(1995)Irwin, Nam, Cabrera, Chugg, and Young]{Irwin1995QET}
K.~D. Irwin, S.~W. Nam, B.~Cabrera, B.~Chugg, and B.~A. Young.
\newblock A quasiparticle-trap-assisted transition-edge sensor for phonon-mediated particle detection.
\newblock \emph{Review of Scientific Instruments}, 66\penalty0 (11):\penalty0 5322--5326, 1995.
\newblock \doi{10.1063/1.1146105}.

\bibitem[Mirabolfathi et~al.(2017)Mirabolfathi, Harris, Mahapatra, Sundqvist, Jastram, Serfass, Faiez, and Sadoulet]{Mirabolfathi2017}
Nader Mirabolfathi, H.~Rusty Harris, Rupak Mahapatra, Kyle Sundqvist, Andrew Jastram, Bruno Serfass, Dana Faiez, and Bernard Sadoulet.
\newblock Toward single electron resolution phonon mediated ionization detectors.
\newblock \emph{Nuclear Instruments and Methods in Physics Research Section A: Accelerators, Spectrometers, Detectors and Associated Equipment}, 855:\penalty0 88--91, 2017.
\newblock \doi{10.1016/j.nima.2017.02.032}.

\bibitem[Anderson et~al.(2011)Anderson, Leman, Doughty, Figueroa-Feliciano, McCarthy, Pyle, and Young]{Anderson2011TES}
A.~J. Anderson, S.~W. Leman, T.~Doughty, E.~Figueroa-Feliciano, K.~A. McCarthy, M.~Pyle, and B.~A. Young.
\newblock Modeling phase-separated transition-edge sensors in {SuperCDMS} detectors, 2011.
\newblock arXiv:1109.3620.

\bibitem[{International Atomic Energy Agency}(2007)]{IAEAAm241}
{International Atomic Energy Agency}.
\newblock Update of {X-ray} and gamma-ray decay data standards for detector calibration and other applications.
\newblock IAEA Nuclear Data Services, 2007.
\newblock $^{241}$Am principal gamma-ray energy: 59.541 keV.

\end{thebibliography}

\end{document}